\pdfoutput=1

\documentclass[aps, prb, reprint, amsmath, amssymb, superscriptaddress]{revtex4-2}
\usepackage[svgnames,psnames]{xcolor}
\usepackage{graphicx}
\usepackage[colorlinks,allcolors = Blue, linktocpage, unicode]{hyperref}

\usepackage{amsmath, amssymb, bm}
\usepackage{braket}
\begin{document}

\title{Thermoelectric properties in semimetals with inelastic electron-hole scattering}

\author{Keigo Takahashi}
\email{takahashi@hosi.phys.s.u-tokyo.ac.jp}
\affiliation{Department of Physics, University of Tokyo, 7-3-1 Hongo, Bunkyo, Tokyo 113-0033, Japan}
\author{Hiroyasu Matsuura}
\affiliation{Department of Physics, University of Tokyo, 7-3-1 Hongo, Bunkyo, Tokyo 113-0033, Japan}
\author{Hideaki Maebashi}
\affiliation{Department of Physics, University of Tokyo, 7-3-1 Hongo, Bunkyo, Tokyo 113-0033, Japan}
\author{Masao Ogata}
\affiliation{Department of Physics, University of Tokyo, 7-3-1 Hongo, Bunkyo, Tokyo 113-0033, Japan}
\affiliation{Trans-Scale Quantum Science Institute, University of Tokyo, 7-3-1 Hongo, Bunkyo, Tokyo 113-0033, Japan} 

\date{\today}

\begin{abstract}
We present systematic theoretical results on thermoelectric effects in semimetals based on the variational method of the linearized Boltzmann equation. Inelastic electron-hole scattering is known to play an important role in the unusual transport of semimetals, including the broad $T^2$ temperature dependence of the electrical resistivity and the strong violation of the Wiedemann-Franz law. By treating the inelastic electron-hole scattering more precisely beyond the relaxation time approximation, we show that the Seebeck coefficient when compensated depends on the screening length of the Coulomb interaction as well as the Lorenz ratio (the ratio of thermal to electric conductivity due to electrons divided by temperature). It is found that deviations from the compensation condition significantly increase the Seebeck coefficient, along with crucial suppressions of the Lorenz ratio. The result indicates that uncompensated semimetals with the electron-hole scattering have high thermoelectric efficiency when the phonon contribution to thermal conductivity is suppressed.
\end{abstract}

\maketitle

\section{Introduction}
Thermoelectric effect or the Seebeck effect, which induces the electromotive force by a temperature gradient, has attracted much attention from the perspective of energy harvesting.
The efficiency of the power generation due to the thermoelectric effect is expressed by a dimensionless figure of merit, $ZT \equiv S^2\sigma T/(\kappa_{\text{el}} + \kappa_{\text{ph}} ) $, where $S$, $\sigma$, $\kappa_{\text{el}}~(\kappa_{\text{ph}})$, and $T$ are the Seebeck coefficient, electrical conductivity, thermal conductivity of electrons~(phonons), and temperature, respectively.
 Materials with large $ZT$ have potential applications in power supplies and thermoelectric cooling. 

Conducting materials can be broadly classified into three categories according to their transport properties: metals, semiconductors, and semimetals \cite{AshcroftMermin}. Metals have the highest electrical conductivity, but they also have proportionally high thermal conductivity and usually satisfy the Wiedemann-Franz (WF) law, which states that the Lorenz ratio ($L = \kappa_{\text{el}}/\sigma T$) becomes the universal constant $L_{0} = \pi^2k_B^2/3e^2$ with $e < 0$ being the charge of an electron. The WF law prevents metals from having large $ZT$. In general, materials that exhibit high thermoelectric performance belong to semiconductors with a large Seebeck coefficient. Thermoelectricity of semimetals, the third category of conducting materials with intermediate conductivity between that of metals and semiconductors, has also been studied for many years~\cite{Ziman2001, Sugihara1969, Thompson1975, Durczewski1991}, and has recently attracted renewed interest~\cite{Beaumale2014, Skinner2018, Markov2019, Han2020, Takahashi2019, Nakano2021}.

The electronic transport due to the electron-hole scattering in semimetals shows several intriguing phenomena, even if the energy dispersion of the model is simple as in Fig.~\ref{dispersion}. First, the electron-hole scattering gives a $T^2$ temperature dependence of the electrical resistivity even without Umklapp process~\cite{Baber1937, Thompson1975, Kukkonen1976, Kukkonen1979, Maldague1979, Oliva1982, Morelli1984}. 
This is because momentum conservation does not necessarily lead to velocity conservation in the case of semimetals. Second, recent experimental and theoretical studies on $\text{WP}_{2}$ have revealed a downward violation of the WF law \cite{Gooth2018, Jaoui2018, Li2018, Zarenia2020, Lee2021}, in which the Lorenz ratio becomes small depending on the screening length of the Coulomb interaction. This is due to the fact that the thermal current is more strongly relaxed than the electrical current due to electron-hole scattering, an effect that goes beyond the relaxation time approximation (RTA) in transport theory. Since the dimensionless figure of merit $ZT$ can be rewritten as
\begin{equation}
ZT = \frac{S^2}{L + \kappa_{\text{ph}}/\sigma T}, \label{ZT_def}
\end{equation}
an unusually small Lorenz ratio in semimetals can lead to a large figure of merit.

In this paper, we systematically study the thermoelectric properties of semimetals using a simple but standard model to clarify the dependences of the electrical, thermal, and thermoelectric transport coefficients on (i) the carrier numbers (compensated, electron-doped, and hole-doped), (ii) the effective masses of electrons and holes, and (iii) the screening length of the Coulomb interaction. In the previous studies, the Lorenz ratio in a compensated semimetal was studied by exact solutions of the Boltzmann equation~\cite{Li2018, Lee2021}. However, this method is not valid for the thermoelectric coefficients. The thermoelectric coefficients due to the electron-hole scattering were studied only for the compensated case by the RTA ~\cite{Lee2021}. Therefore, the general behavior of thermoelectric coefficients for the uncompensated semimetal with the electron-hole scattering is unclear. 
In addition, the RTA is not exact for inelastic scattering~\cite{AshcroftMermin} and the importance of inelastic scattering in a semimetal has been discussed \cite{Takahashi2019}. Therefore,
it should be testified whether the RTA is valid or not by the analysis beyond RTA. Analysis by the trial functions is useful to consider transports in the presence of the inelastic scattering and employed in various systems, such as graphene and bilayer graphene~\cite{Zarenia2019_MG, Zarenia2019_BLG, Nguyen2020}. 
Here, we apply the variational method~\cite{Ziman2001} to the linearized Boltzmann equation, which is more reliable than RTA. We will show that there is a contribution to the thermoelectric effect that is not captured by the RTA in the previous study. In the present paper, we focus on the effect of the electron-hole scattering, and the effect of phonons is out of the scope of this paper \footnote{Phonons may give two contributions. One is the electron-phonon scattering. This can be incorporated by adding a scattering term. The other is the thermal conductivity of phonons $\kappa_{\text{ph}}$, which we neglect in the estimation of the figure of merit in this paper.}. In the following, we study the temperature range $k_BT/\Delta \lesssim  0.06$ where $\Delta$ is an energy offset (see Fig.\ref{dispersion}) since the electron-hole scattering becomes dominant in low-temperature region compared to the electron-phonon scattering.

This paper is organized as follows. In Sec. \ref{Sec:Model_and_Boltzmann}, we introduce the model and the Boltzmann equation. In particular, in Sec. \ref{Subsec:Model}, we provide a detailed description of our model, illustrated in Fig. \ref{dispersion}. In Sec. \ref{Subsec:Boltzmann}, we introduce a systematic method based on the Boltzmann equation to calculate the transport coefficients for this model. The results and discussions are given in Sec. \ref{Sec:Results_Discussions}. First, we present the temperature dependence of transport coefficients. Then, we discuss the carrier-number dependence of thermoelectric properties when the electron-hole scattering dominates. 
In the compensated case, the Seebeck coefficient is zero if the effective masses of the electrons and holes are the same. If the effective masses are different the Seebeck coefficient becomes finite, but small. However, we will show that it is sensitive to the screening length of the Coulomb interaction as is the case for the Lorenz ratio. In the uncompensated cases, we find that slight deviations from the compensation bring a large Seebeck coefficient when the electron-hole scattering dominates. We also estimate $\widetilde{ZT} \equiv S^2\sigma T/\kappa_{\text{el}} = S^2/L$,  which gives an upper bound of the figure of merit,  in our framework, and find that the electron-hole scattering gives large $\widetilde{ZT}$ in the uncompensated case due to the collaboration of the reduction of the Lorenz ratio and the increase of the Seebeck coefficient.
Finally, the conclusions are given in Sec. \ref{Sec:Conclusion}.

\section{Model and Boltzmann equation}\label{Sec:Model_and_Boltzmann}
\subsection{Model}\label{Subsec:Model}
\begin{figure}[tbp]
\begin{center}
\rotatebox{0}{\includegraphics[angle=0,width=1\linewidth]{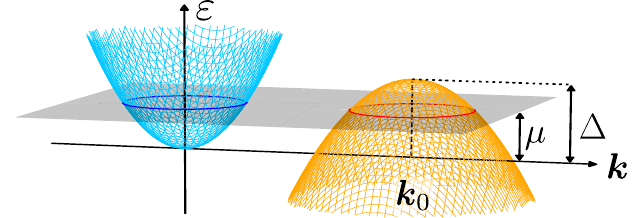}}
\caption{Two-band model consisting of electron (blue) and hole (orange) bands.}
\label{dispersion}
\end{center}
\end{figure}
We study a two-band model depicted in Fig.~\ref{dispersion} consisting of electron and hole bands with three-dimensional quadratic dispersions~\cite{Li2018, Lee2021}
\begin{equation}
\varepsilon_{1,\bm{k}} = \frac{\hbar^2\bm{k}^2}{2m_{1}}, \varepsilon_{2,\bm{k}} = \Delta - \frac{\hbar^2 (\bm{k} - \bm{k}_0)^2}{2m_{2}}
\end{equation}
where $m_{1}$ ($m_{2}$) is the effective mass of electrons (holes), and $\Delta$ is the energy offset. Therefore, both carriers have spherical Fermi surfaces. 

The number of electrons (holes) is given by
$n_{1} = V^{-1}\sum_{\bm{k}}  2 f_0(\varepsilon_{1,\bm{k}})$ ($n_{2}  =  V^{-1}  \sum_{\bm{k}} 2(1 - f_0(\varepsilon_{2,\bm{k}}) )$)
where a factor 2 and $V$ indicate the spin degeneracy, and the volume of the system, and $f_{0}(\varepsilon) = (e^{\beta(\varepsilon - \mu)} + 1)^{-1}$ is the Fermi-Dirac distribution function with $\beta = (k_BT)^{-1}$ and $\mu$ is the chemical potential which keeps the net charge $e\Delta n = e(n_1 - n_2)$ at the value of $T = 0$.
By introducing a parameter $\chi$ defined by $k_{\text{F},2} = \chi k_{\text{F},1}$, we obtain $n_{2} = \chi^3 n_{1}$ at $T = 0$ and the Fermi energy ($\varepsilon_{\text{F}}$) is given by $\varepsilon_{\text{F}} = m_{2}\Delta/(\chi^{2} m_{1} +  m_{2})$.  
As a typical scale of wavenumber, we define $k_{\text{F}} = \sqrt{2m_{1}m_{2}\Delta/\hbar^2(m_{1} + m_{2})}$, which is the Fermi wavenumber in the case of $\chi = 1$, which corresponds to the compensated case, $n_{1} = n_{2}$.
\subsection{Boltzmann equation and variational method}\label{Subsec:Boltzmann}
The Boltzmann equation of the system is given by~\cite{Ziman2001, Li2018, Lee2021}
\begin{eqnarray}
&&\left[- eE_x v^{(l)}_{\bm{k};x} - (\varepsilon_{l,\bm{k}} - \mu)v^{(l)}_{\bm{k};x} \left(- \frac{\nabla_x T}{T} \right) \right] \left(- \frac{\partial f_0(\varepsilon_{l,\bm{k}})}{\partial \varepsilon_{l,\bm{k}}} \right) \nonumber \\
&& = \left(\frac{\partial f^{(l)}(\bm{k})}{\partial t} \right)_{\text{imp}} + \left(\frac{\partial f^{(l)}(\bm{k})}{\partial t} \right)_{\text{e-h}} + \left(\frac{\partial f^{(l)}(\bm{k})}{\partial t} \right)_{\text{e-e}} \label{boltmznn} 
\end{eqnarray}
where $v^{(l)}_{\bm{k};x} = \hbar^{-1 }\nabla_{k_x} \varepsilon_{l,\bm{k}}~(l = 1,2)$ is the velocity of the band $l$~($l = 1,2$).
$E_{x}$ and $(- \nabla_{x} T/T)$ are the electric field and the temperature gradient along the $x$ axis, respectively. The three terms on the right-hand side of eq.~(\ref{boltmznn}) represent the impurity, interband (electron-hole), and intraband (electron-electron and hole-hole) scattering, respectively. We assume that the impurity scattering is due to the short-range impurity potential and the inter- and intraband scattering are due to the screened Coulomb interaction where we neglect the exchange process~\cite{Li2018, Lee2021}. For example, the electron-hole scattering for the band $l = 1$ is given by ~\cite{Li2018, Lee2021}
\begin{widetext}
\begin{eqnarray}
\left( \frac{\partial f^{(1)}(\bm{k})}{\partial t} \right)_{\text{e-h}} 
=&&-  2\sum_{\bm{k}_2,\bm{k}_3,\bm{k}_4}  S_{\text{e-h}}(\bm{k},\bm{k}_2;\bm{k}_3,\bm{k}_4) f^{(1)}(\bm{k})f^{(2)}(\bm{k}_2) (1- f^{(1)}(\bm{k}_3)) (1- f^{(2)}(\bm{k}_4))  \nonumber \\
&&+ 2\sum_{\bm{k}_1,\bm{k}_2,\bm{k}_4}  S_{\text{e-h}}(\bm{k}_1,\bm{k}_2;\bm{k},\bm{k}_4) f^{(1)}(\bm{k}_1)f^{(2)}(\bm{k}_2) (1- f^{(1)}(\bm{k})) (1- f^{(2)}(\bm{k}_4)) ,
\end{eqnarray}
where the factor 2 is the spin degeneracy and $S_{\text{e-h}}(\bm{k}_1,\bm{k}_2;\bm{k}_3,\bm{k}_4)$ is given by
\begin{equation}
S_{\text{e-h}}(\bm{k}_1,\bm{k}_2;\bm{k}_3,\bm{k}_4) = \frac{2\pi}{\hbar} \frac{1}{V^2} \left(\frac{1}{4\pi \varepsilon_{0}} \right)^2 \left(\frac{4\pi e^2}{|\bm{k}_1 - \bm{k}_3|^2 + \alpha^2} \right)^2 \frac{(2\pi)^3}{V} \delta( \bm{k}_1 + \bm{k}_2 - \bm{k}_3 - \bm{k}_4) \delta(\varepsilon_{1,\bm{k}_1} + \varepsilon_{2,\bm{k}_2} - \varepsilon_{1,\bm{k}_3} - \varepsilon_{2,\bm{k}_4}). \label{screened_coulomb}
\end{equation}
\end{widetext}
Here, $\varepsilon_{0}$ is the dielectric constant and $\alpha$ represents the inverse of the Thomas-Fermi screening length, where $\alpha^2 = e^2( m_{1} k_{\text{F},1} + m_{2} k_{\text{F},2})/\pi^2 \hbar^2\varepsilon_{0}$~\cite{Li2018, Lee2021}. The other scattering terms have similar forms, which are given in Appendix \ref{appendix:scattering_terms}.

In the variational method~\cite{Ziman2001}, the distribution function is expanded as $f^{(l)}(\bm{k}) = f_{0}(\varepsilon_{l,\bm{k}}) + \beta f_{0}(\varepsilon_{l,\bm{k}})(1 - f_{0}(\varepsilon_{l,\bm{k}})) \Phi^{(l)}(\bm{k})$
where $\Phi^{(l)}$ is assumed to be small. Keeping terms up to the first order of $\Phi^{(l)}$, eq.~(\ref{boltmznn}) can be rewritten as~\cite{Ziman2001}
\begin{equation}
X^{(l)} = P^{(l)}[\Phi] = P_{\text{imp}}^{(l)}[\Phi] + P_{\text{e-h}}^{(l)}[\Phi] + P_{\text{e-e}}^{(l)}[\Phi],
\end{equation}
where $-X^{(l)}$ denotes the left hand side of eq.~(\ref{boltmznn}). Note that $P_{\text{imp}}^{(l)}[\Phi]$ and $ P_{\text{e-e}}^{(l)}[\Phi]$ are linear functionals of $\Phi^{(l)}$, while $P_{\text{e-h}}^{(l)}[\Phi]$ is a linear functional of $(\Phi^{(1)},\Phi^{(2)})$. Explicit forms of these scattering terms are presented in Appendix \ref{appendix:scattering_terms}.
Then, we assume that the trial function for the band $l$ is given by
$ \Phi^{(l)}(\bm{k}) = \sum_{i = 1}^{2} \eta_{i}^{(l)} \varphi_{i}^{(l)}(\bm{k}), \label{tri_Func} $
where $\varphi_{i}^{(l)}(\bm{k}) = v^{(l)}_{\bm{k};x}(\varepsilon_{l,\bm{k}} - \mu)^{i-1}$ and the coefficients $\eta_{i}^{(l)}$ are determined so as to maximize a variational functional. Using the variational method~\cite{Ziman2001,SM}, we obtain the expression for $\eta_{i}^{(l)}$ as 
\begin{equation}
\eta_{i}^{(l)} =\sum_{j,k} (P^{-1})^{(lk)}_{ij}\left[ J^{(k)}_{j} E_x + U^{(k)}_{j} \left(- \frac{\nabla_x T}{T} \right) \right],
\end{equation}
where $P^{(lk)}_{ij}$ is a matrix representation of $P^{(l)}[\Phi]$ for the chosen basis $\{\varphi_{i}^{(l)}(\bm{k})\}$ and 
\begin{equation}
\left(\begin{array}{c}
J_{i}^{(l)} \\
U_{i}^{(l)}  \\
    \end{array}
\right) = \frac{1}{V} \sum_{\bm{k}} \varphi_{i}^{(l)}(\bm{k}) v_{\bm{k};x}^{(l)} \left(\begin{array}{c}
  e \\
  \varepsilon_{l,\bm{k}} - \mu  \\
      \end{array}
  \right)\left( - \frac{\partial f_{0}(\varepsilon_{l,\bm{k}})}{\partial \varepsilon_{l,\bm{k}}} \right).
\end{equation}
The explicit form of matrix $P^{(lk)}_{ij}$ is given in Appendix \ref{appendix:matrix_P}. Since $P^{(l)}_{\text{e-h}}[\Phi]$ contains the distribution function of the other band, we have the superscript $(lk)$ in $P_{ij}^{(lk)}$. 
In the evaluation of $P_{ij}^{(lk)}$, we analytically perform angular integrals, and numerically evaluate the remaining energy integrals. The details are given in the Supplemental Material (SM)~\cite{SM}. 
Transport coefficients ($L_{11}$, $L_{12} = L_{21}$, and $L_{22}$), which relate the electric (heat) current $J_{x}~(J_{q;x})$ to the external fields, defined as
\begin{eqnarray}
\left(\begin{array}{c}
    J_{x} \\
    J_{q;x}  \\
    \end{array}
\right) 
= 
\left(\begin{array}{cc}
    L_{11} & L_{12}\\
    L_{21} & L_{22}\\
    \end{array}
\right)\left(\begin{array}{c}
    E_x \\
    - \nabla_x T/T \\
    \end{array}
\right),
\end{eqnarray}
are given by
\begin{equation}
\left(\begin{array}{cc}
L_{11} & L_{12}\\
L_{21} & L_{22}\\
    \end{array}
\right) = 2\sum_{i,l,j,k} 
\left(\begin{array}{c}
J^{(l)}_{i}\\
U^{(l)}_{i}\\
    \end{array}
\right) (P^{-1})^{(lk)}_{ij} (J_{j}^{(k)},U_{j}^{(k)})  .
\end{equation}
In the degenerate regime ($k_BT \ll \varepsilon_{\text{F}}$), these transport coefficients are approximated as
\begin{eqnarray}
L_{11} &\simeq& 2({}^{t} \bm{J}_1 P_{11}^{-1}\bm{J}_1), \label{L11_approx} \\
L_{12} &\simeq& 2[{}^{t} \bm{J}_1 P_{11}^{-1}\bm{U}_1 + ({}^{t}\bm{J}_2 - {}^{t}\bm{J}_1 P_{11}^{-1}P_{12}) P_{22}^{-1}\bm{U}_2 ], \label{L12_approx} \\
L_{22} &\simeq& 2({}^{t} \bm{U}_1 P_{11}^{-1}\bm{U}_1 + {}^{t}\bm{U}_2 P_{22}^{-1}\bm{U}_2 ), \label{L22_approx}
\end{eqnarray}
by considering the power of $k_BT/\varepsilon_{\text{F}}$ where $\bm{J}_i = {}^{t} (J^{(1)}_i,J^{(2)}_i)$ and $\bm{U}_i = {}^{t} (U^{(1)}_i,U^{(2)}_i)$.
It should be noted that although ${}^{t} \bm{U}_1 P_{11}^{-1}\bm{U}_1$ in $L_{22}$ is not the leading order, we consider this term because it corresponds to the ambipolar contribution~\cite{Ziman2001, Zarenia2020, Lee2021}.

\section{Results and discussions}\label{Sec:Results_Discussions}

\begin{figure}[tbp]
\begin{center}
\rotatebox{0}{\includegraphics[angle=0,width=1\linewidth]{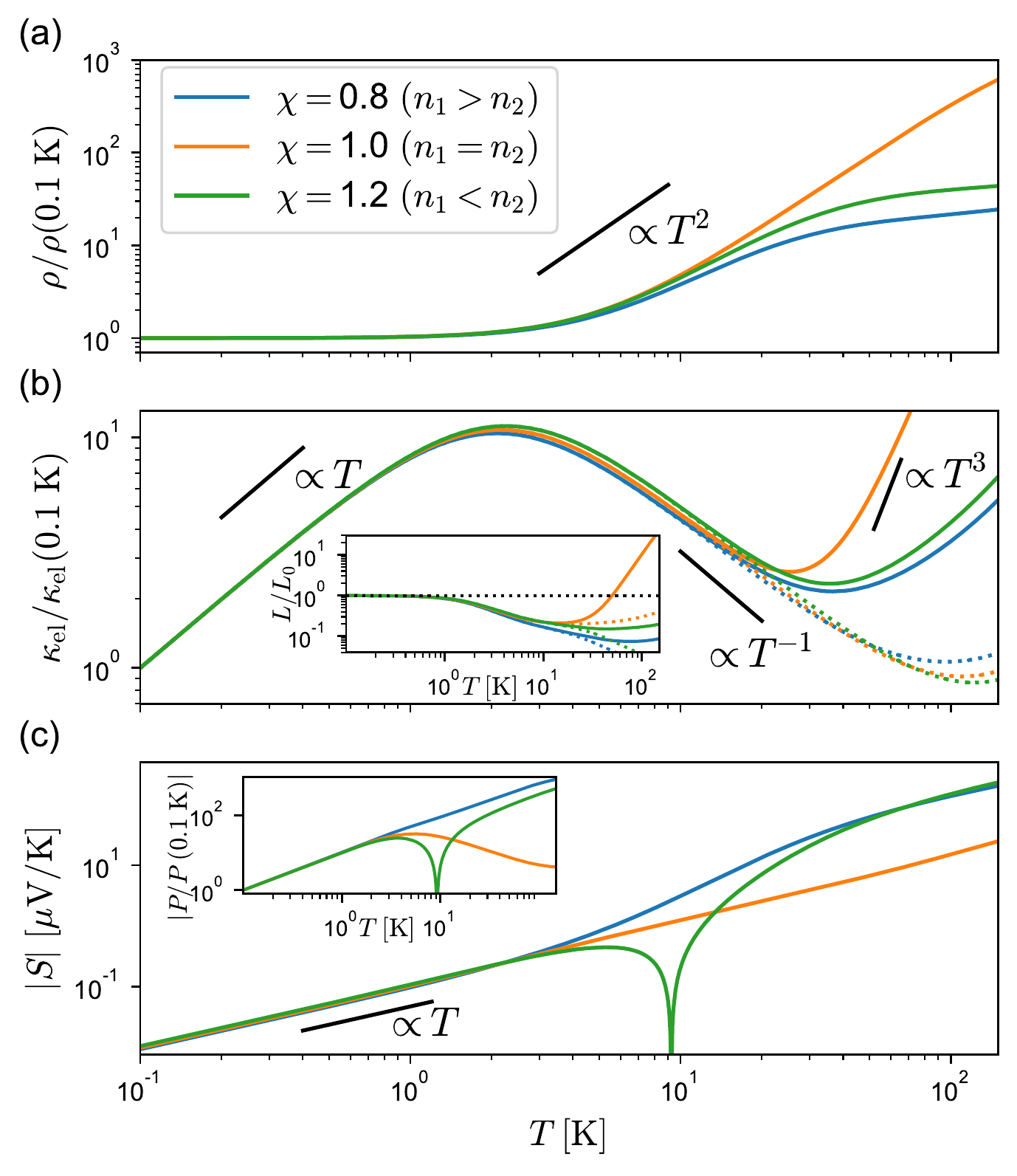}}
\caption{Temperature dependences of (a) electrical resistivity, (b) thermal conductivity, and (c) the absolute value of Seebeck coefficient for three values of $\chi$ with $m_{2} = 3m_{1} = 3m_{\text{e}}$. We normalize the electrical resistivity and thermal conductivity by the value at $T = 0.1$ K. The dotted lines in (b) represent $\tilde{\kappa}_{\text{el}}$ defined in the text. In (c), the Seebeck coefficient for $\chi = 1.2$ changes its sign at $T = 10$ K. The inset of (b) shows the temperature dependence of the normalized Lorenz ratio $L/L_0 = \kappa_{\text{el}}\rho/TL_0$ (solid lines) and $\tilde{\kappa}_{\text{el}}\rho/TL_0$ (dotted lines) with $L_0=\pi^2k_B^2/3 e^2$. The inset of (c) indicates temperature dependence of the Peltier conductivity $P \equiv S\sigma$.}
\label{el_th_se}
\end{center}
\end{figure}

\subsection{Temperature dependence}
Figure~\ref{el_th_se} shows the temperature dependences of resistivity $\rho = \sigma^{-1} = L_{11}^{-1}$, thermal conductivity $\kappa_{\text{el}} = (L_{22} - L_{21}L_{12}/L_{11})/T$, and the Seebeck coefficient $S = L_{12}/TL_{11}$ for three values of $\chi$.  Lorenz ratio and Peltier conductivity $P \equiv S\sigma$ are also shown in the inset of (b) and (c), respectively.  Here, we set $m_{2} =  3m_{1} =  3m_{\text{e}}$ with $m_{\text{e}}$ being the electron mass and $\varepsilon_{0}$ so that $\alpha = k_{\text{F}}$ at $\chi = 1$. In the following, we use $\Delta = 0.2~\text{eV}$ as a typical value. The strength of the impurity scattering is chosen so that the electron-hole scattering dominates above 4 K (see Appendix \ref{appendix:scattering_terms} for the choices of parameters).

\subsubsection{Electrical resistivity}
The electrical resistivity $\rho$ (Fig.~\ref{el_th_se}(a)) is independent of $T$ in the region of $T \lesssim 4~\text{K}$, because the impurity scattering dominates. As temperature increases, $\rho$ shows $T^2$-dependence due to the electron-hole scattering ($4 ~\text{K} \lesssim T \lesssim 30~\text{K}$). On the other hand, in high temperature regime ($T \gtrsim 30~\text{K}$), $\rho$ in the compensated case ($\chi = 1.0$) shows $T^2$-dependence, while $\rho$ saturates in the uncompensated cases. This is because the contribution to the electric current from the total momentum, which is relaxed only through momentum dissipative scatterings, is proportional to $n_1 - n_2$, and this contribution does not vanish in the uncompensated case \cite{Kukkonen1976} (see also SM \cite{SM}). When the system is uncompensated and the relative momentum is strongly relaxed by the electron-hole scattering, the relaxation of the total momentum by the impurity scattering governs the electric conduction~\cite{Kukkonen1976, Kukkonen1979, Pal2012}. Then, the saturated resistivity $\rho_{\text{sat}}$ obeys $\rho_{\text{sat}}/\rho(T = 0) \sim [(n_1 + n_2)/(n_1 - n_2)]^2$ \cite{Kukkonen1976, SM}.

\subsubsection{Thermal conductivity}
The thermal conductivity $\kappa_{\text{el}}$ (Fig.~\ref{el_th_se}(b)) is proportional to $T$ at low temperatures. As shown in the inset, WF law holds in this temperature region. In the intermediate temperature region, $\kappa_{\text{el}}$ decreases slightly slower than to $T^{-1}$. This temperature dependence is approximately consistent with WF law ($\kappa_{\text{el}}\propto T\sigma$), but the normalized Lorenz ratio is less than 1 and temperature dependent. 

For the compensated case ($\chi = 1$), this result is consistent with previous studies~\cite{Li2018,Zarenia2020,Lee2021}. In particular, $T^3$ dependence is due to the ambipolar effect~\cite{Zarenia2020, Lee2021}.
Actually, $\tilde{\kappa}_{\text{el}} \equiv 2({}^{t}\bm{U}_2 P_{22}^{-1}\bm{U}_2 )/T$ (dotted lines in Fig.~\ref{el_th_se}(b)), which does not include the ambipolar contribution, does not show the increase but instead has the $T^{-1}$-dependence in a wider range of temperatures. For $T > 60~\mathrm{K}$, $\tilde{\kappa}_{\text{el}}$ upwardly deviates from $T^{-1}$  due to the subleading temperature dependence~\cite{SM}. 

In contrast, in the uncompensated case, $\kappa_{\text{el}}$ does not follow $T^{3}$-dependence and is smaller than $\kappa_{\text{el}}$ for $\chi = 1$. As a result, $L/L_0$ (inset) is small even in the high temperature region. This means that the ambipolar contribution is not large when uncompensated.
The ambipolar contribution is associated with the transport of the compensated electrons and holes moving in the same direction under the temperature gradient giving no electric current as discussed for semiconductors \cite{Ziman2001}. This ambipolar contribution is also present in semimetals~\cite{Zarenia2020, Lee2021} and is weak in the uncompensated case. We can show \cite{SM} that, in $\kappa_{\text{el}} = (L_{22} - L_{21}L_{12}/L_{11})/T$, the enhancement of $L_{12}$ in the uncompensated case cancels the ambipolar contribution in $L_{22}$ leading to the suppression of the Lorenz ratio. 
\subsubsection{Seebeck coefficient}
The Seebeck coefficient $S$ is negative and almost independent of $\chi$ in the low temperature region. This is because the transport property is mainly determined by the electrons that have a smaller effective mass than the holes. 
In this low temperature region, the impurity scattering is dominant and thus the Mott formula is valid. At higher temperatures, $S$ for $\chi=1.2$ changes its sign to positive at $T \sim 10$ K. This can be understood as follows. In the high temperature region, the relative momentum between electrons and holes is strongly relaxed by the electron-hole scattering, and thus the electric current is mainly carried by the total momentum proportional to $e(n_1 - n_2)$~\cite{Kukkonen1976}. As a result, the sign of $S$ is determined by the holes for the case of $\chi = 1.2$ $(n_2 > n_1)$. In the temperature region above 40 K, $S$ becomes again almost linear in $T$. Apparently, the coefficient of the linear $T$ term for $\chi \neq 1$ is about ten times larger than that at low temperatures. 

\begin{figure}[tbp]
\begin{center}
\rotatebox{0}{\includegraphics[angle=0,width=1\linewidth]{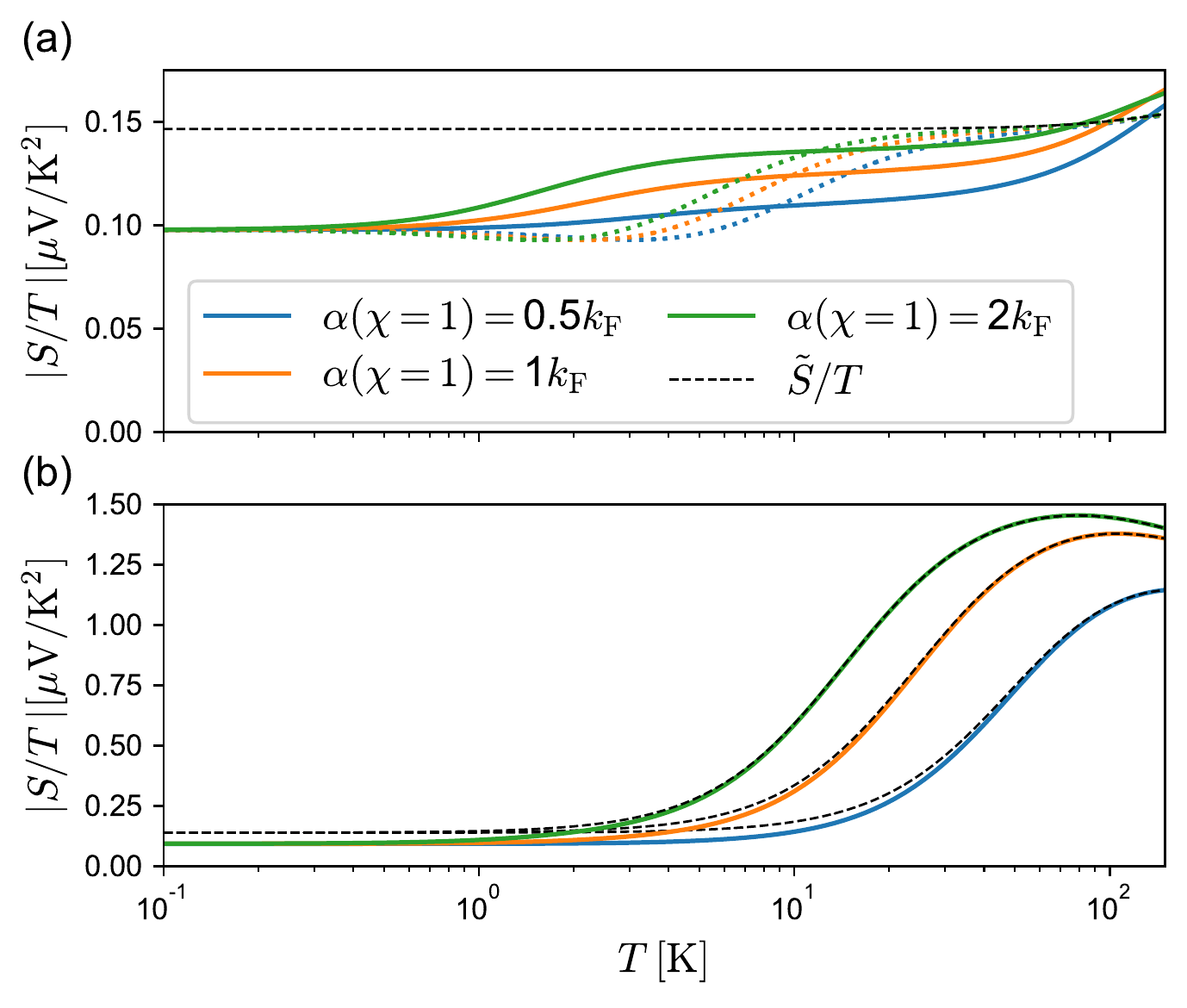}}
\caption{Temperature dependences of $|S/T|$ for three screening lengths, or equivalently three dielectric constants for (a) $\chi = 1$ and (b) $\chi = 0.8$. The dielectric constants are chosen so that $\alpha = 0.5 k_{\text{F}}, 1.0k_{\text{F}},$ and $2.0k_{\text{F}}$ at $\chi =1$. Colored dotted lines in (a) show results when considering the impurity and intraband scattering. }
\label{se_alpha}
\end{center}
\end{figure} 

First, let us study the Seebeck coefficient for the compensated case ($\chi=1$) more closely. We plot in Fig.~\ref{se_alpha} the temperature dependence of $|S/T|$ for three screening lengths. We can see that the coefficient of the linear-$T$ term gradually increases as a function of $T$ and reaches some value, which depends on $\alpha$.
The black dashed line in Fig.~\ref{se_alpha} (a) shows $\tilde{S}/T \equiv {}^{t} \bm{J}_1 P_{11}^{-1}\bm{U}_1/T^2({}^{t}\bm{J}_1 P_{11}^{-1}\bm{J}_1)$, which is equivalent to the RTA as shown in the SM \cite{SM}. Apparently, $\tilde S/T$ does not depend on $\alpha$, which is consistent with the previous study \cite{Lee2021} showing that the Seebeck coefficient in the RTA does not depend on the relaxation time by the electron-hole scattering (see also SM \cite{SM}). This indicates that the RTA does not explain the $\alpha$ dependence of $S$ in Fig.~\ref{se_alpha} for the compensated case. 
To see the effect of the electron-hole scattering, we show the results (colored dotted lines in Fig.~\ref{se_alpha}) in which the electron-hole scattering is neglected and only the impurity and intraband scattering are considered. In this case, $|S/T|$ in the high temperature region does not depend on $\alpha$, which means that the interband scattering plays an important role in $\alpha$ dependence of $S$. Since the first term of the right-hand side of eq.~(\ref{L12_approx}) corresponds to the RTA~\cite{SM}, the $\alpha$ dependence comes from the other terms in eq.~ (\ref{L12_approx}), i.e., from the terms including $P_{12}$ and $P_{22}$. In particular, $P_{12}$ is nonzero for the electron-hole scattering unlike the intraband scattering.

For the uncompensated cases ($\chi \neq 1$), the situation is different. Figure~\ref{se_alpha}(b) shows the temperature dependence of $|S/T|$ for three screening lengths in the case of $\chi = 0.8$. In this case, the enhancement of $|S/T|$ at high temperatures is larger than that for $\chi = 1$. In the uncompensated cases, ${}^{t} \bm{J}_1 P_{11}^{-1}\bm{U}_1$ in eq.~(\ref{L12_approx}) gives the major contribution when the electron-hole scattering dominates~\cite{SM}. As a result, $S$ can be approximated as $\tilde{S} = {}^{t} \bm{J}_1 P_{11}^{-1}\bm{U}_1/T({}^{t}\bm{J}_1 P_{11}^{-1}\bm{J}_1)$, which is equivalent to the RTA. In fact, $\tilde{S}/T$, which are shown in the black dashed lines in Fig.~\ref{se_alpha}(b), reproduce $S/T$ at high temperatures.

\subsection{Carrier-number dependence}
\begin{figure}[tbp]
\begin{center}
\rotatebox{0}{\includegraphics[angle=0,width=1\linewidth]{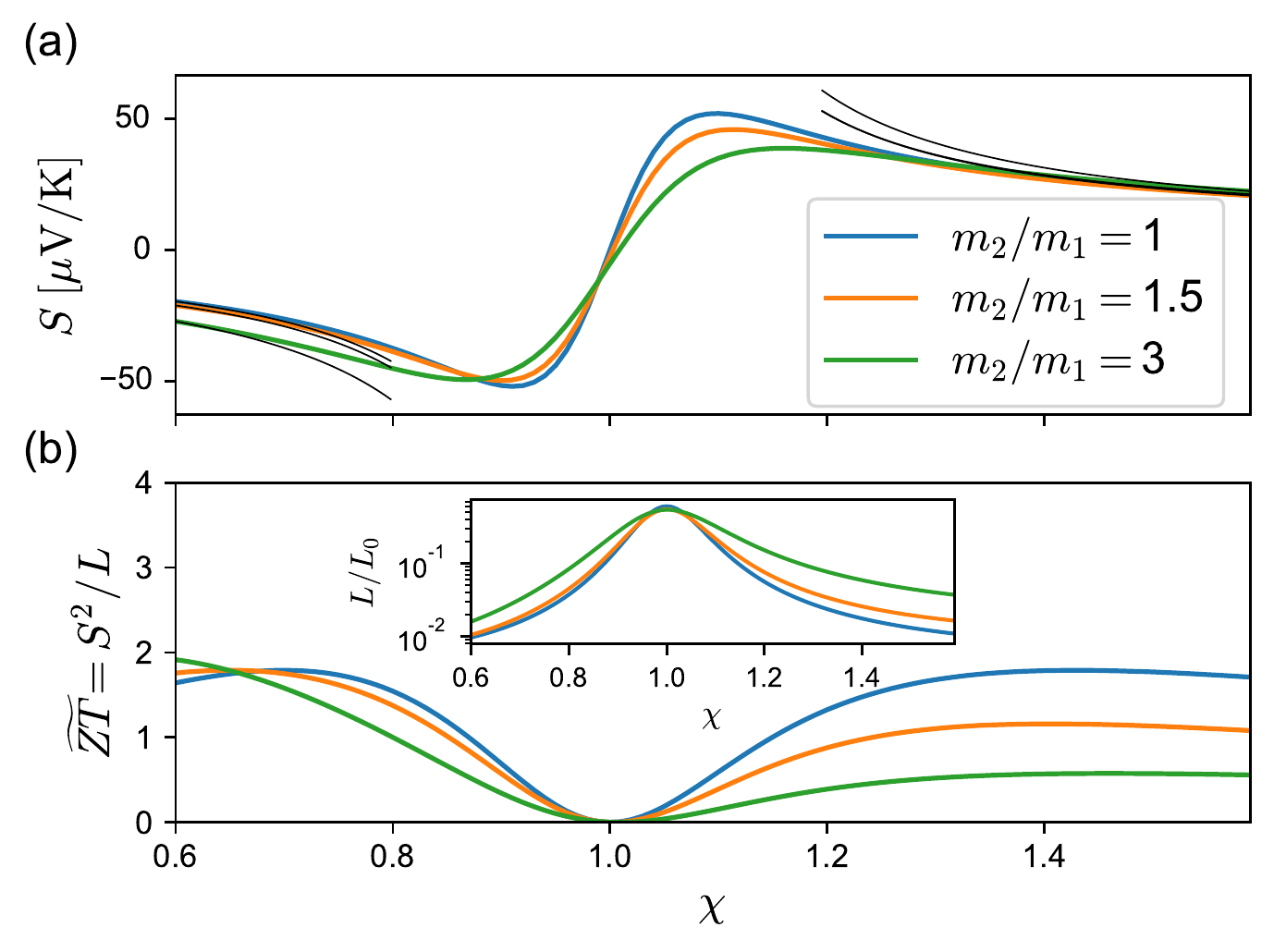}}
\caption{$\chi$ dependences of (a) Seebeck coefficient and (b) $\widetilde{ZT} = S^2\sigma T/\kappa_{\text{el}}=S^2/L$ at $40 ~\text{K}$. $\varepsilon_{0}$ is the same as in Fig.~\ref{el_th_se}. The inset of (b) shows the Lorenz ratio. 
Black lines in (a) are proportional to $(m_{1} + \chi m_{2})(\chi^2 m_{1} + m_{2})/(\chi^3 - 1)$.} 
\label{chi_dep}
\end{center}
\end{figure}
To understand the effect of doping and the difference in effective masses, we plot in Fig.~\ref{chi_dep} the $\chi$ dependence of the Seebeck coefficient, the Lorenz ratio (inset), and $\widetilde{ZT} = S^2\sigma T/\kappa_{\text{el}}= S^2/L$ for three values of $m_{2}/m_{1}$ at $T=40 ~\text{K}$.
We can see that the absolute value of $S$ increases and the Lorenz ratio drastically decreases in the uncompensated case. As a result, $\widetilde{ZT}$ (Fig.~\ref{chi_dep}(b)) drastically increases in the uncompensated case. For the case of $m_{2}/m_{1}=1$, $\widetilde{ZT}$ is almost symmetric with respect to $\chi$, while $\widetilde{ZT}$ is larger for the case with $\chi<1$, i.e., $n_{1} > n_{2}$, than for the case with $\chi>1$ when $m_{2} / m_{1}>1$. As we can see from Fig.~\ref{el_th_se}(c), $S$ at $T=40 ~\text{K}$ is almost the same for $\chi=0.8$ and $\chi=1.2$. Thus, the difference in $\widetilde{ZT}$ comes from the difference in the Lorenz ratio. 

In the limit of vanishing the impurity scattering, the Seebeck coefficient behaves as $S \simeq \tilde{S} \sim (m_{1} + \chi m_{2})(\chi^2 m_{1} + m_{2})/(\chi^3 - 1) \propto (n_{2} - n_{1})^{-1}$~\cite{SM}. We see this dependence in the black lines of Fig.~\ref{chi_dep}(a) for $\chi$ far away from 1. This means that a slight deviation from compensation can lead to a large Seebeck coefficient. As discussed in Ref.~\cite{Kukkonen1979} in connection to the Hall coefficient, we can interpret the behavior of $S$ as follows: the electrons and holes are locked by the electron-hole scattering and they can be treated as a single carrier with charge $e(n_{1} - n_{2})$. Note the present result is similar to that of the carrier-number dependence of the Seebeck coefficient in graphene and bilayer graphene~\cite{Zarenia2019_BLG, Zarenia2019_MG}.
Controlling carrier numbers to make a small deviation from compensation is a strategy for the large Seebeck coefficient as observed in $\mathrm{Ti}_{1 + x} \mathrm{S}_2$~\cite{Beaumale2014,Thompson1975}. Our results suggest that this strategy in clean semimetals with dominating electron-hole scattering is also beneficial for reducing the Lorenz ratio and then achieving high $ZT$.

\section{Conclusions}\label{Sec:Conclusion}
In the present paper, we studied the effects of electron-hole scattering with a finite screening length within the Boltzmann transport theory. However, when we consider the Kubo-Luttinger linear response theory in the case of the finite-range Coulomb interaction, there is an additional contribution in the heat current operator that does not satisfy the Sommerfeld-Bethe relation~\cite{Ogata2019, Takarada2021}. Since this type of heat current operator is not taken into account in the Boltzmann equation, the microscopic study based on the Kubo-Luttinger formalism might reveal a new aspect of transport in semimetals.

In conclusion, we have studied transport coefficients of semimetals considering the impurity, electron-hole, and intraband scattering based on the variational analysis of the Boltzmann equation. We have shown that the thermoelectric coefficient of semimetals with the electron-hole scattering contains contributions beyond the RTA. The part neglected in the RTA brings the screening lengths dependence of the Seebeck coefficient. We have also shown that when the electron-hole scattering dominates in the uncompensated cases the Seebeck coefficient is largely enhanced. $\widetilde{ZT}$, in which the phononic thermal conductivity is neglected, can be large in the uncompensated condition due to the increase of the Seebeck coefficient and the reduction of the Lorenz ratio. Although our analysis does not take into account $\kappa_{\text{ph}}$, this semimetal system can be a very good candidate for thermoelectric devices. 

\section*{Acknowledgments}
We are grateful to Prof. Akitoshi Nakano for fruitful discussions. This work is supported by Grants-in-Aid for Scientific Research from the Japan Society for the Promotion of Science (Nos. JP22K18954, JP20K03802, JP21K03426, and JP18K03482), and JST-Mirai Program Grant (No. JPMJMI19A1). K. T. is supported by Forefront Physics and Mathematics Program to Drive Transformation (FoPM).

\appendix 
\begin{widetext}
\section{Scattering terms} \label{appendix:scattering_terms}
The scattering terms are presented. The impurity scattering is given by
\begin{equation}
\left( \frac{\partial f^{(l)}(\bm{k})}{\partial t} \right)_{\text{imp}} = - \sum_{\bm{k}'} S_{\text{imp}}^{(l)}(\bm{k},\bm{k}')f^{(l)}(\bm{k})(1- f^{(l)}(\bm{k}')) 
 + \sum_{\bm{k}'} S_{\text{imp}}^{(l)}(\bm{k}',\bm{k})f^{(l)}(\bm{k}')(1- f^{(l)}(\bm{k})),
\end{equation}
where $l$ represents the band $l = 1,2$, and
\begin{equation}
S_{\text{imp}}^{(l)}(\bm{k}',\bm{k}) = S_0 \delta(\varepsilon_{l,\bm{k}} - \varepsilon_{l,\bm{k}'}).
\end{equation}
$S_0$ determines the strength of the impurity scattering. We set $V S_0 \tilde{k}_{\text{F}}^3/4\pi^2 = 1.35 \times 10^{-11}~\text{J/s}$, where $\tilde{k}_{\text{F}}$ is a Fermi wavenumber for $m_{2} = 3m_{1} = 3m_{\mathrm{e}}$, $\Delta = 0.2~\text{eV}$ and $\chi =1$.

The electron-hole scattering for the band $l = 2$ is given by
\begin{eqnarray}
\left( \frac{\partial f^{(2)}(\bm{k})}{\partial t} \right)_{\text{e-h}} 
& = & -  2\sum_{\bm{k}_1,\bm{k}_3,\bm{k}_4}  S_{\text{e-h}}(\bm{k}_1,\bm{k};\bm{k}_3,\bm{k}_4) f^{(1)}(\bm{k}_1)f^{(2)}(\bm{k}) (1- f^{(1)}(\bm{k}_3)) (1- f^{(2)}(\bm{k}_4)) \nonumber \\
&&+ 2\sum_{\bm{k}_1,\bm{k}_2,\bm{k}_3}  S_{\text{e-h}}(\bm{k}_1,\bm{k}_2;\bm{k}_3,\bm{k}) f^{(1)}(\bm{k}_1)f^{(2)}(\bm{k}_2) (1- f^{(1)}(\bm{k}_3)) (1- f^{(2)}(\bm{k})),
\end{eqnarray}

The intraband scattering for the band $l$ is given by
\begin{eqnarray}
\left( \frac{\partial f^{(l)}(\bm{k})}{\partial t} \right)_{\text{e-e}} 
&=& -  2\sum_{\bm{k}_2,\bm{k}_3,\bm{k}_4}  S^{(l)}_{\text{e-e}}(\bm{k},\bm{k}_2;\bm{k}_3,\bm{k}_4) f^{(l)}(\bm{k})f^{(l)}(\bm{k}_2) (1- f^{(l)}(\bm{k}_3)) (1- f^{(l)}(\bm{k}_4))   \nonumber \\
&&+ 2\sum_{\bm{k}_1,\bm{k}_2,\bm{k}_4}  S^{(l)}_{\text{e-e}}(\bm{k}_1,\bm{k}_2;\bm{k},\bm{k}_4) f^{(l)}(\bm{k}_1)f^{(l)}(\bm{k}_2) (1- f^{(l)}(\bm{k})) (1- f^{(l)}(\bm{k}_4)),
\end{eqnarray}
and 
\begin{equation}
S^{(l)}_{\text{e-e}}(\bm{k}_1,\bm{k}_2;\bm{k}_3,\bm{k}_4) = \frac{2\pi}{\hbar} \frac{1}{V^2} \left(\frac{1}{4\pi \varepsilon_{0}} \right)^2 \left(\frac{4\pi e^2}{|\bm{k}_1 - \bm{k}_3|^2 + \alpha^2} \right)^2 \frac{(2\pi)^3}{V} \delta( \bm{k}_1 + \bm{k}_2 - \bm{k}_3 - \bm{k}_4) \delta(\varepsilon_{l,\bm{k}_1} + \varepsilon_{l,\bm{k}_2} - \varepsilon_{l,\bm{k}_3} - \varepsilon_{l,\bm{k}_4}),
\end{equation}
where we use the same screened Coulomb potential as in $S_{\text{e-h}}(\bm{k}_1,\bm{k}_2;\bm{k}_3,\bm{k}_4)$.

The linearized forms of scattering terms are given by
\begin{equation}
P_{\text{imp}}^{(l)}[\Phi] = \frac{1}{k_B T}\sum_{\bm{k}'} S_{\text{imp}}^{(l)}(\bm{k}',\bm{k})f_{0}(\varepsilon_{l,\bm{k}})(1- f_{0}(\varepsilon_{l,\bm{k}'}))(\Phi^{(l)}(\bm{k}) - \Phi^{(l)}(\bm{k}')) ,
\end{equation}
\begin{eqnarray}
P_{\text{e-h}}^{(1)}[\Phi]
=  \frac{2}{k_B T} \sum_{\bm{k}_2,\bm{k}_3,\bm{k}_4} &&  S_{\text{e-h}}(\bm{k}_1,\bm{k}_2;\bm{k}_3,\bm{k}_4)  f_{0}(\varepsilon_{1,\bm{k}_1 }) f_{0}(\varepsilon_{2,\bm{k}_2}) (1- f_{0}(\varepsilon_{1,\bm{k}_3})) (1- f_{0}(\varepsilon_{2,\bm{k}_4}))  \nonumber \\
&& \times (\Phi^{(1)}(\bm{k}_1) + \Phi^{(2)}(\bm{k}_2) - \Phi^{(1)}(\bm{k}_3) - \Phi^{(2)}(\bm{k}_4)) ,\\
P_{\text{e-h}}^{(2)}[\Phi]
=\frac{2}{k_B T}\sum_{\bm{k}_1,\bm{k}_3,\bm{k}_4} && S_{\text{e-h}}(\bm{k}_1,\bm{k}_2;\bm{k}_3,\bm{k}_4) f_{0}(\varepsilon_{1,\bm{k}_1 }) f_{0}(\varepsilon_{2,\bm{k}_2}) (1- f_{0}(\varepsilon_{1,\bm{k}_3}))  (1- f_{0}(\varepsilon_{2,\bm{k}_4}))  \nonumber \\
&& \times (\Phi^{(1)}(\bm{k}_1) + \Phi^{(2)}(\bm{k}_2) - \Phi^{(1)}(\bm{k}_3) - \Phi^{(2)}(\bm{k}_4)) ,\\
P_{\text{e-e}}^{(l)}[\Phi]
= \frac{2}{k_B T}\sum_{\bm{k}_2,\bm{k}_3,\bm{k}_4} && S^{(l)}_{\text{e-e}}(\bm{k}_1,\bm{k}_2;\bm{k}_3,\bm{k}_4) f_{0}(\varepsilon_{l,\bm{k}_1 }) f_{0}(\varepsilon_{l,\bm{k}_2}) (1- f_{0}(\varepsilon_{l,\bm{k}_3}))  (1- f_{0}(\varepsilon_{l,\bm{k}_4}))  \nonumber \\
&& \times (\Phi^{(l)}(\bm{k}_1) + \Phi^{(l)}(\bm{k}_2) - \Phi^{(l)}(\bm{k}_3) - \Phi^{(l)}(\bm{k}_4)).
\end{eqnarray}

\section{Matrix $P^{(lk)}_{ij}$} \label{appendix:matrix_P}
The matrix $P^{(lk)}_{ij}$ is understood as a matrix representation of the scatterings. This is given by $P^{(lk)}_{ij} = P^{(lk)}_{\text{imp},ij} + P^{(lk)}_{\text{e-h},ij} + P^{(lk)}_{\text{e-e},ij}$ where
\begin{eqnarray}
P^{(lk)}_{\text{imp},ij} &=& \frac{ \delta_{lk} }{2k_BT V} \sum_{\bm{k},\bm{k}'} S_{\text{imp}}^{(l)}(\bm{k},\bm{k}')f_{0}(\varepsilon_{l,\bm{k}})(1- f_{0}(\varepsilon_{l,\bm{k}'}))(\varphi^{(l)}_i (\bm{k}) -\varphi^{(l)}_i  (\bm{k}'))(\varphi^{(l)}_j  (\bm{k}) -\varphi^{(l)}_j  (\bm{k}')), \label{scat_P_imp} \\
P^{(11)}_{\text{e-h},ij} 
&=& \frac{1}{k_B TV}\sum_{\bm{k}_1,\bm{k}_2,\bm{k}_3,\bm{k}_4} S_{\text{e-h}}(\bm{k}_1,\bm{k}_2;\bm{k}_3,\bm{k}_4) 
f_{0}(\varepsilon_{1,\bm{k}_1}) f_{0}(\varepsilon_{2,\bm{k}_2}) (1- f_{0}(\varepsilon_{1,\bm{k}_3})) (1- f_{0}(\varepsilon_{2,\bm{k}_4})) \nonumber \\ 
&& \times (\varphi^{(1)}_{i}(\bm{k}_{1}) - \varphi^{(1)}_{i}(\bm{k}_{3})) (\varphi^{(1)}_{j}(\bm{k}_{1}) - \varphi^{(1)}_{j}(\bm{k}_{3})), \label{scat_P_eh^11}\\
P^{(12)}_{\text{e-h},ij} = P^{(21)}_{\text{e-h},ji}
&=& \frac{1}{k_B TV}\sum_{\bm{k}_1,\bm{k}_2,\bm{k}_3,\bm{k}_4} S_{\text{e-h}}(\bm{k}_1,\bm{k}_2;\bm{k}_3,\bm{k}_4) 
f_{0}(\varepsilon_{1,\bm{k}_1}) f_{0}(\varepsilon_{2,\bm{k}_2}) (1- f_{0}(\varepsilon_{1,\bm{k}_3})) (1- f_{0}(\varepsilon_{2,\bm{k}_4})) \nonumber \\ 
&& \times (\varphi^{(1)}_{i}(\bm{k}_{1}) - \varphi^{(1)}_{i}(\bm{k}_{3})) (\varphi^{(2)}_{j}(\bm{k}_{2}) - \varphi^{(2)}_{j}(\bm{k}_{4})), \label{scat_P_eh^12}\\
P^{(22)}_{\text{e-h},ij} 
&=& \frac{1}{k_B TV}\sum_{\bm{k}_1,\bm{k}_2,\bm{k}_3,\bm{k}_4} S_{\text{e-h}}(\bm{k}_1,\bm{k}_2;\bm{k}_3,\bm{k}_4) 
f_{0}(\varepsilon_{1,\bm{k}_1}) f_{0}(\varepsilon_{2,\bm{k}_2}) (1- f_{0}(\varepsilon_{1,\bm{k}_3})) (1- f_{0}(\varepsilon_{2,\bm{k}_4})) \nonumber \\ 
&& \times (\varphi^{(2)}_{i}(\bm{k}_{2}) - \varphi^{(2)}_{i}(\bm{k}_{4})) (\varphi^{(2)}_{j}(\bm{k}_{2}) - \varphi^{(2)}_{j}(\bm{k}_{4})), \label{scat_P_eh^22}\\
P^{(lk)}_{\text{e-e},ij} 
&=& \frac{\delta_{lk}}{2k_B TV}\sum_{\bm{k}_1,\bm{k}_2,\bm{k}_3,\bm{k}_4} S^{(l)}_{\text{e-e}}(\bm{k}_1,\bm{k}_2;\bm{k}_3,\bm{k}_4) 
f_{0}(\varepsilon_{l,\bm{k}_1}) f_{0}(\varepsilon_{l,\bm{k}_2}) (1- f_{0}(\varepsilon_{l,\bm{k}_3})) (1- f_{0}(\varepsilon_{l,\bm{k}_4})) \nonumber \\ 
&& \times (\varphi^{(l)}_{i}(\bm{k}_{1}) + \varphi^{(l)}_{i}(\bm{k}_{2}) - \varphi^{(l)}_{i}(\bm{k}_{3}) - \varphi^{(l)}_{i}(\bm{k}_{4})) (\varphi^{(l)}_{j}(\bm{k}_{1}) + \varphi^{(l)}_{j}(\bm{k}_{2}) - \varphi^{(l)}_{j}(\bm{k}_{3}) - \varphi^{(l)}_{j}(\bm{k}_{4})). \label{scat_P_ee}
\end{eqnarray}
\end{widetext}

\bibliographystyle{apsrev4-2}
\bibliography{main}

\end{document}


\title{Supplemental Material for \\"Thermoelectric properties in semimetals with inelastic electron-hole scattering"}

\author{Keigo Takahashi}
\email{takahashi@hosi.phys.s.u-tokyo.ac.jp}
\affiliation{Department of Physics, University of Tokyo, 7-3-1 Hongo, Bunkyo, Tokyo 113-0033, Japan}
\author{Hiroyasu Matsuura}
\affiliation{Department of Physics, University of Tokyo, 7-3-1 Hongo, Bunkyo, Tokyo 113-0033, Japan}
\author{Hideaki Maebashi}
\affiliation{Department of Physics, University of Tokyo, 7-3-1 Hongo, Bunkyo, Tokyo 113-0033, Japan}
\author{Masao Ogata}
\affiliation{Department of Physics, University of Tokyo, 7-3-1 Hongo, Bunkyo, Tokyo 113-0033, Japan}
\affiliation{Trans-Scale Quantum Science Institute, University of Tokyo, 7-3-1 Hongo, Bunkyo, Tokyo 113-0033, Japan} 

\date{\today}
\maketitle

\tableofcontents

\setcounter{section}{0}
\setcounter{equation}{0}
\setcounter{figure}{0}
\setcounter{table}{0}
\setcounter{page}{1}
\makeatletter
\renewcommand{\thesection}{\Alph{section}}
\renewcommand{\thesubsection}{\roman{subsection}}
\renewcommand{\theequation}{\Alph{section}\arabic{equation}}
\renewcommand{\thefigure}{\Alph{section}\arabic{figure}}
\renewcommand{\thetable}{\Alph{section}\arabic{table}}

\setcounter{equation}{0}
\setcounter{figure}{0}
\setcounter{table}{0}
\section{Temperature dependence of chemical potential}
For a parameter $\chi$ and temperature $T$, the chemical potential $\mu$ is determined so as to meet the following equation:
\begin{equation}
\Delta n = \frac{k_{\text{F,1}}^3}{3\pi^2} - \frac{k_{\text{F,2}}^3}{3\pi^2} 
= \int  2D_1(\varepsilon) f_0(\varepsilon) d\varepsilon - \int 2D_2(\varepsilon)(1 - f_0(\varepsilon) ) d\varepsilon.
\end{equation}
Here, $D_{l}(\varepsilon)$ denotes the density of states for each band per spin, given by
\begin{equation}
D_{1}(\varepsilon) = \frac{1}{4\pi^2} \left( \frac{2m_{1}}{\hbar^2} \right)^{3/2} \sqrt{\varepsilon} \Theta(\varepsilon), \qquad D_{2}(\varepsilon) = \frac{1}{4\pi^2} \left( \frac{2m_{2}}{\hbar^2} \right)^{3/2} \sqrt{\Delta - \varepsilon}\Theta(\Delta - \varepsilon),
\end{equation}
where the $\Theta(x)$ is the Heaviside step function. Within the parameters and temperature considered in this paper, $0 < \mu < \Delta$ is satisfied.

\setcounter{equation}{0}
\setcounter{figure}{0}
\setcounter{table}{0}
\section{Variational method}
\subsection{Proof of variational principle}
In this section, following Ziman's argument~\cite{Ziman2001}, we show that the true solution of the Boltzmann equation is given by the function that maximizes the variation functional under a constraint.

Let us assume that $\tilde{\Phi}^{(l)}(\bm{k})$ is the true solution of the Boltzmann equation satisfying
\begin{equation}
  X^{(l)} = P^{(l)}[\tilde{\Phi}] = P_{\text{imp}}^{(l)}[\tilde{\Phi}] + P_{\text{e-h}}^{(l)}[\tilde{\Phi}] + P_{\text{e-e}}^{(l)}[\tilde{\Phi}]. \label{lin_Boltzmann}
\end{equation}
Then, we consider a trial function $\Phi^{(l)}(\bm{k})$ which satisfies the constraint $\sum_{l} \braket{\Phi^{(l)},X^{(l)}} = \sum_{l} \braket{\Phi^{(l)},P^{(l)}[\Phi]}$, where $\braket{A,B} = V^{-1}\sum_{\bm{k}} A(\bm{k})B(\bm{k}) $. 
Let us define a functional $\mathcal{F}$ as
\begin{equation}
\mathcal{F}[\varphi,\tilde{\varphi}] =\sum_{l} \braket{\varphi^{(l)},P^{(l)}[\tilde{\varphi}]},
\end{equation}
and $f[\varphi] = \mathcal{F}[\varphi,\varphi] $. Substituting $\Phi - \tilde{\Phi}$ into $f$, and using the non-negative nature and $\mathcal{F}[\Phi , \tilde{\Phi}] = \mathcal{F}[\tilde{\Phi},\Phi]$, we obtain
\begin{equation}
f[\Phi - \tilde{\Phi}] = \mathcal{F}[\Phi - \tilde{\Phi},\Phi - \tilde{\Phi}] = \mathcal{F}[\Phi ,\Phi ] - 2\mathcal{F}[\Phi , \tilde{\Phi}]  + \mathcal{F}[\tilde{\Phi},\tilde{\Phi}] \geq 0. \label{eneq_f}
\end{equation}
By using eq.~(\ref{lin_Boltzmann}) and the above constraint for $\Phi^{(l)}(\bm{k})$, we find
\begin{eqnarray}
\mathcal{F}[\Phi , \tilde{\Phi}] 
&=&\sum_{l} \braket{\Phi^{(l)},P^{(l)}[\tilde{\Phi}]} \nonumber \\
&=&\sum_{l} \braket{\Phi^{(l)},X^{(l)}} \nonumber \\
&=&\sum_{l} \braket{\Phi^{(l)},P^{(l)}[\Phi]} = \mathcal{F}[\Phi , \Phi].
\end{eqnarray}
Therefore, eq.~(\ref{eneq_f}) becomes $- \mathcal{F}[\Phi ,\Phi ] + \mathcal{F}[\tilde{\Phi},\tilde{\Phi}] \geq 0$, which leads to
\begin{equation}
f[\tilde{\Phi}] = \mathcal{F}[\tilde{\Phi},\tilde{\Phi}]  \geq \mathcal{F}[\Phi ,\Phi ] = f[\Phi ].
\end{equation}
This means that the true solution of the Boltzmann equation is the argument of the maximum of the variational functional $f[\Phi]$ under the constraint.

\subsection{Derivation of coefficients $\eta^{(l)}_i$}
We can find the expression of $\eta^{(l)}_i$ by the method of Lagrange multiplier. 
We maximize the functional $h[\Phi]$ defined as
\begin{equation}
h[\Phi]
= f\left[\Phi\right] -\lambda \sum_{l}\left[\braket{\Phi^{(l)},P^{(l)}[\Phi]}  - \braket{\Phi^{(l)},X^{(l)}} \right],
\end{equation}
where $\lambda$ is the Lagrange multiplier.
Using the trial function introduced in the main text, the functional $f\left[\Phi\right]$ becomes the quadratic form of $\eta^{(l)}_{i}$ as
\begin{equation}
f\left[\Phi\right] = \sum_{i,j} \sum_{l,k} \eta^{(l)}_{i}\eta^{(k)}_j P^{(lk)}_{ij},
\end{equation}
where $P^{(lk)}_{ij} = P^{(lk)}_{\text{imp},ij} + P^{(lk)}_{\text{e-h},ij} + P^{(lk)}_{\text{e-e},ij}$ and 
\begin{eqnarray}
P^{(lk)}_{\text{imp},ij} &=& \frac{ \delta_{lk} }{2k_BT V} \sum_{\bm{k},\bm{k}'} S_{\text{imp}}^{(l)}(\bm{k},\bm{k}')f_{0}(\varepsilon_{l,\bm{k}})(1- f_{0}(\varepsilon_{l,\bm{k}'}))(\varphi^{(l)}_i (\bm{k}) -\varphi^{(l)}_i  (\bm{k}'))(\varphi^{(l)}_j  (\bm{k}) -\varphi^{(l)}_j  (\bm{k}')), \label{scat_P_imp} \\
P^{(11)}_{\text{e-h},ij} 
&=& \frac{1}{k_B TV}\sum_{\bm{k}_1,\bm{k}_2,\bm{k}_3,\bm{k}_4} S_{\text{e-h}}(\bm{k}_1,\bm{k}_2;\bm{k}_3,\bm{k}_4) 
f_{0}(\varepsilon_{1,\bm{k}_1}) f_{0}(\varepsilon_{2,\bm{k}_2}) (1- f_{0}(\varepsilon_{1,\bm{k}_3})) (1- f_{0}(\varepsilon_{2,\bm{k}_4})) \nonumber \\ 
&& \times (\varphi^{(1)}_{i}(\bm{k}_{1}) - \varphi^{(1)}_{i}(\bm{k}_{3})) (\varphi^{(1)}_{j}(\bm{k}_{1}) - \varphi^{(1)}_{j}(\bm{k}_{3})), \label{scat_P_eh^11}\\
P^{(12)}_{\text{e-h},ij} = P^{(21)}_{\text{e-h},ji}
&=& \frac{1}{k_B TV}\sum_{\bm{k}_1,\bm{k}_2,\bm{k}_3,\bm{k}_4} S_{\text{e-h}}(\bm{k}_1,\bm{k}_2;\bm{k}_3,\bm{k}_4) 
f_{0}(\varepsilon_{1,\bm{k}_1}) f_{0}(\varepsilon_{2,\bm{k}_2}) (1- f_{0}(\varepsilon_{1,\bm{k}_3})) (1- f_{0}(\varepsilon_{2,\bm{k}_4})) \nonumber \\ 
&& \times (\varphi^{(1)}_{i}(\bm{k}_{1}) - \varphi^{(1)}_{i}(\bm{k}_{3})) (\varphi^{(2)}_{j}(\bm{k}_{2}) - \varphi^{(2)}_{j}(\bm{k}_{4})), \label{scat_P_eh^12}\\
P^{(22)}_{\text{e-h},ij} 
&=& \frac{1}{k_B TV}\sum_{\bm{k}_1,\bm{k}_2,\bm{k}_3,\bm{k}_4} S_{\text{e-h}}(\bm{k}_1,\bm{k}_2;\bm{k}_3,\bm{k}_4) 
f_{0}(\varepsilon_{1,\bm{k}_1}) f_{0}(\varepsilon_{2,\bm{k}_2}) (1- f_{0}(\varepsilon_{1,\bm{k}_3})) (1- f_{0}(\varepsilon_{2,\bm{k}_4})) \nonumber \\ 
&& \times (\varphi^{(2)}_{i}(\bm{k}_{2}) - \varphi^{(2)}_{i}(\bm{k}_{4})) (\varphi^{(2)}_{j}(\bm{k}_{2}) - \varphi^{(2)}_{j}(\bm{k}_{4})), \label{scat_P_eh^22}\\
P^{(lk)}_{\text{e-e},ij} 
&=& \frac{\delta_{lk}}{2k_B TV}\sum_{\bm{k}_1,\bm{k}_2,\bm{k}_3,\bm{k}_4} S^{(l)}_{\text{e-e}}(\bm{k}_1,\bm{k}_2;\bm{k}_3,\bm{k}_4) 
f_{0}(\varepsilon_{l,\bm{k}_1}) f_{0}(\varepsilon_{l,\bm{k}_2}) (1- f_{0}(\varepsilon_{l,\bm{k}_3})) (1- f_{0}(\varepsilon_{l,\bm{k}_4})) \nonumber \\ 
&& \times (\varphi^{(l)}_{i}(\bm{k}_{1}) + \varphi^{(l)}_{i}(\bm{k}_{2}) - \varphi^{(l)}_{i}(\bm{k}_{3}) - \varphi^{(l)}_{i}(\bm{k}_{4})) (\varphi^{(l)}_{j}(\bm{k}_{1}) + \varphi^{(l)}_{j}(\bm{k}_{2}) - \varphi^{(l)}_{j}(\bm{k}_{3}) - \varphi^{(l)}_{j}(\bm{k}_{4})). \label{scat_P_ee}
\end{eqnarray}

Detailed calculations of these matrix elements are discussed later. From the above expression, the functional $h[\Phi]$ becomes a function of 
\begin{equation}
h(\eta_{i}^{(l)},\lambda) = (1 - \lambda) \sum_{i,j} \sum_{l,k} \eta^{(l)}_{i}\eta^{(k)}_j P^{(lk)}_{ij} + \lambda \sum_{i} \sum_{l}  \eta_{i}^{(l)} \left[ J^{(l)}_{i} E_x + U^{(l)}_{i} \left(- \frac{\nabla_x T}{T} \right) \right].
\end{equation}

The solutions of $\partial h(\eta_{i}^{(l)},\lambda)/\partial \eta_{i}^{(l)} = 0$ and $\partial h(\eta_{i}^{(l)},\lambda)/\partial \lambda = 0$ give the maximum value of $h$. Therefore, we have to solve
\begin{eqnarray}
\frac{\partial h(\eta_{i}^{(l)},\lambda)}{\partial \eta_{i}^{(l)}} &=& 2(1 - \lambda) \sum_{j} \sum_{k} P^{(lk)}_{ij} \eta_{j}^{(k)} + \lambda  \left[ J^{(l)}_{i} E_x + U^{(l)}_{i} \left(- \frac{\nabla_x T}{T} \right) \right] = 0, \label{df_deta} \\
\frac{\partial h(\eta_{i}^{(l)},\lambda)}{\partial \lambda} &=& - \sum_{i,j} \sum_{l,k} \eta^{(l)}_{i}\eta^{(k)}_j P^{(lk)}_{ij} + \sum_{i} \sum_{l}  \eta_{i}^{(l)} \left[ J^{(l)}_{i} E_x + U^{(l)}_{i} \left(- \frac{\nabla_x T}{T} \right) \right] = 0.
\end{eqnarray}
With a little calculation, we reach 
\begin{equation}
(2 - \lambda) \sum_{i,j} \sum_{l,k} \eta^{(l)}_{i}\eta^{(k)}_j P^{(lk)}_{ij} = (2 - \lambda) f[\Phi]  = 0.
\end{equation}
Hence $f[\Phi] > 0$, we obtain $\lambda = 2$. This means that eq.~(\ref{df_deta}) leads to
\begin{equation}
\sum_{j,k} P^{(lk)}_{ij} \eta_{j}^{(k)} =  J^{(l)}_{i} E_x + U^{(l)}_{i} \left(- \frac{\nabla_x T}{T} \right),
\end{equation}
and thus
\begin{equation}
\eta_{i}^{(l)} = \sum_{j,k} (P^{-1})^{(lk)}_{ij}\left[ J^{(k)}_{j} E_x + U^{(k)}_{j} \left(- \frac{\nabla_x T}{T} \right) \right].
\end{equation}

\subsection{Expressions of transport coefficients}
We can calculate electric current $J_x$ and heat current $J_{q;x}$ with the expression of $\eta_{i}^{(l)}$. Taking the spin degrees of freedom into account, we obtain
\begin{eqnarray}
J_{x} 
&=& \frac{2e}{V}\sum_{l} \sum_{\bm{k}} v^{(l)}_{\bm{k};x} f^{(l)}(\bm{k}) \nonumber \\
&=& \frac{2e}{V}\sum_{l} \sum_{\bm{k}} v^{(l)}_{\bm{k};x}\Phi^{(l)}(\bm{k})\left(- \frac{\partial f_{0}(\varepsilon_{l,\bm{k}})}{\partial \varepsilon_{l,\bm{k}}} \right) \nonumber \\
&=& 2\sum_{i,l} \eta^{(l)}_{i}J_{i}^{(l)} \nonumber \\
&=& 2\sum_{i,l} \sum_{j,k} (P^{-1})^{(lk)}_{ij}\left[ J^{(k)}_{j} E_x + U^{(k)}_{j} \left(- \frac{\nabla_x T}{T} \right) \right] J_{i}^{(l)} \nonumber \\
&=& 2\sum_{i,l} \sum_{j,k}  J_{i}^{(l)}  (P^{-1})^{(lk)}_{ij} J^{(k)}_{j} E_x +  2\sum_{i,l} \sum_{j,k} J_{i}^{(l)}  (P^{-1})^{(lk)}_{ij}  U^{(k)}_{j} \left(- \frac{\nabla_x T}{T} \right).
\end{eqnarray}
Therefore,
\begin{eqnarray}
L_{11} &=& 2\sum_{i,l} \sum_{j,k}  J_{i}^{(l)}  (P^{-1})^{(lk)}_{ij} J^{(k)}_{j}, \\
L_{12} &=& 2\sum_{i,l} \sum_{j,k}  J_{i}^{(l)}  (P^{-1})^{(lk)}_{ij} U^{(k)}_{j}.
\end{eqnarray}

The calculation for the heat current is parallel to that for the electric current:
\begin{eqnarray}
J_{q;x} 
&=& \frac{2}{V}\sum_{l} \sum_{\bm{k}} (\varepsilon_{l,\bm{k}} - \mu)v^{(l)}_{\bm{k};x} f^{(l)}(\bm{k}) \nonumber \\
&=& 2\sum_{i,l} \sum_{j,k}  U_{i}^{(l)}  (P^{-1})^{(lk)}_{ij} J^{(k)}_{j} E_x +  2\sum_{i,l} \sum_{j,k} U_{i}^{(l)}  (P^{-1})^{(lk)}_{ij}  U^{(k)}_{j} \left(- \frac{\nabla_x T}{T} \right).
\end{eqnarray}
Therefore,
\begin{equation}
L_{22} = 2 \sum_{i,l} \sum_{j,k}  U_{i}^{(l)}  (P^{-1})^{(lk)}_{ij} U^{(k)}_{j}.
\end{equation}

\setcounter{equation}{0}
\setcounter{figure}{0}
\setcounter{table}{0}
\section{Calculation of $J^{(l)}_{i}$, $U^{(l)}_{i}$, and $P_{ij}^{(lk)}$}
In this section, we calculate $J^{(l)}_{i}$, $U^{(l)}_{i}$, and $P_{ij}^{(lk)}$. Here, the momentum of holes is measured from $\bm{k}_{0}$.

\subsection{$J^{(l)}_{i}$ and $U^{(l)}_{i}$}
$J^{(l)}_{1}$ is calculated as 
\begin{eqnarray}
J_{i}^{(1)} 
&=& \frac{e}{V} \sum_{\bm{k}} \varphi_{i}^{(1)}(\bm{k}) v_{\bm{k};x}^{(1)} \left( - \frac{\partial f_{0}(\varepsilon_{1,\bm{k}})}{\partial \varepsilon_{1,\bm{k}}} \right) \nonumber \\
&=&  \frac{e}{V} \sum_{\bm{k}} (v_{\bm{k};x}^{(1)})^2 (\varepsilon_{1,\bm{k}} - \mu)^{i-1} \left( - \frac{\partial f_{0}(\varepsilon_{1,\bm{k}})}{\partial \varepsilon_{1,\bm{k}}} \right) \nonumber \\
&=& \frac{2e}{3m_{1}V} \sum_{\bm{k}} \varepsilon_{1,\bm{k}} (\varepsilon_{1,\bm{k}} - \mu)^{i-1} \left( - \frac{\partial f_{0}(\varepsilon_{1,\bm{k}})}{\partial \varepsilon_{1,\bm{k}}} \right) \nonumber \\
&=& \frac{2e}{3m_{1}} \int d\varepsilon D_1(\varepsilon) \varepsilon (\varepsilon - \mu)^{i-1} \left( - \frac{\partial f_{0}(\varepsilon)}{\partial \varepsilon} \right).
\end{eqnarray}
The calculation of $J^{(l)}_{2}$ is the same as $J^{(l)}_{1}$,
\begin{eqnarray}
J_{i}^{(2)} 
&=& \frac{e}{V} \sum_{\bm{k}} \varphi_{i}^{(2)}(\bm{k}) v_{\bm{k};x}^{(2)} \left( - \frac{\partial f_{0}(\varepsilon_{2,\bm{k}})}{\partial \varepsilon_{2,\bm{k}}} \right) \nonumber \\
&=& \frac{e}{V} \sum_{\bm{k}} (v_{\bm{k};x}^{(2)})^2 (\varepsilon_{2,\bm{k}} - \mu)^{i-1} \left( - \frac{\partial f_{0}(\varepsilon_{2,\bm{k}})}{\partial \varepsilon_{2,\bm{k}}} \right) \nonumber \\
&=& \frac{2e}{3m_{2}V} \sum_{\bm{k}} (\Delta - \varepsilon_{2,\bm{k}}) (\varepsilon_{2,\bm{k}} - \mu)^{i-1} \left( - \frac{\partial f_{0}(\varepsilon_{2,\bm{k}})}{\partial \varepsilon_{2,\bm{k}}} \right) \nonumber \\
&=& \frac{2e}{3m_{2}} \int d\varepsilon D_2(\varepsilon)(\Delta - \varepsilon) (\varepsilon - \mu)^{i-1} \left( - \frac{\partial f_{0}(\varepsilon)}{\partial \varepsilon} \right).
\end{eqnarray}
$U_{i}^{(l)} $ is computed in a parallel way, 
\begin{eqnarray}
U_{i}^{(1)} 
&=& \frac{2}{3m_{1}} \int d\varepsilon D_1(\varepsilon) \varepsilon (\varepsilon - \mu)^{i} \left( - \frac{\partial f_{0}(\varepsilon)}{\partial \varepsilon} \right), \\
U_{i}^{(2)} 
&=& \frac{2}{3m_{2}} \int d\varepsilon D_2(\varepsilon)(\Delta - \varepsilon) (\varepsilon - \mu)^{i} \left( - \frac{\partial f_{0}(\varepsilon)}{\partial \varepsilon} \right).
\end{eqnarray}

We can analytically calculate $J_{1}^{(1)}$ as
\begin{eqnarray}
J_{1}^{(1)} 
&=&\frac{2e}{3m_{1}} \int d\varepsilon D_{1}(\varepsilon) \varepsilon \left(- \frac{\partial f_0(\varepsilon)}{\partial \varepsilon} \right) \nonumber \\
&=&\frac{e}{m_{1}} \int d\varepsilon D_{1}(\varepsilon) f_0(\varepsilon) \nonumber \\
&=& \frac{e}{m_{1}} D_1(\mu) \mu \left( \frac{k_BT}{\mu} \right)^{3/2} \frac{\sqrt{\pi}}{2} (- \text{Li}_{\frac{3}{2}}(- e^{\beta \mu})),
\end{eqnarray}
where $\text{Li}_{n}(x)$ is the polylogarithm function.
The other terms can be also expressed in this way as follows:
\begin{eqnarray}
J_{2}^{(1)} &=&  e U_{1}^{(1)} =
\frac{e}{m_{1}} D_1(\mu) \mu  \left( \frac{k_BT}{\mu} \right)^{3/2} \left[- \frac{5\sqrt{\pi}}{4} k_BT 
\text{Li}_{\frac{5}{2}}(- e^{\beta \mu})  +   \frac{\sqrt{\pi}}{2} \mu  \text{Li}_{\frac{3}{2}}(- e^{\beta \mu})  \right], \\
U_{2}^{(1)} 
&=&  \frac{1}{m_{1}} D_1(\mu) \mu  \left( \frac{k_BT}{\mu} \right)^{3/2} \nonumber \\
&& \times \left[
- \frac{35\sqrt{\pi}}{8} (k_BT)^{2}  \text{Li}_{\frac{7}{2}}(- e^{\beta \mu})  
+ \frac{5\sqrt{\pi}}{2}  k_BT \mu  \text{Li}_{\frac{5}{2}}(- e^{\beta \mu}) 
- \frac{\sqrt{\pi}}{2} \mu^2 \text{Li}_{\frac{3}{2}}(- e^{\beta \mu})  \right],
\end{eqnarray}
and 
\begin{eqnarray}
J_{1}^{(2)} 
&=&  \frac{e}{m_{2}} D_2(\mu) (\Delta - \mu) \left( \frac{k_BT}{\Delta - \mu} \right)^{3/2} \frac{\sqrt{\pi}}{2} (- \text{Li}_{\frac{3}{2}}(- e^{\beta (\Delta - \mu)})), \\
J_{2}^{(2)} &=&  e U_{1}^{(2)} =
\frac{e}{m_{2}} D_2(\mu) (\Delta - \mu) \left( \frac{k_BT}{\Delta - \mu} \right)^{3/2} \left[ \frac{5\sqrt{\pi}}{4} k_BT 
\text{Li}_{\frac{5}{2}}(- e^{\beta (\Delta - \mu)})  -  \frac{\sqrt{\pi}}{2}(\Delta - \mu) \text{Li}_{\frac{3}{2}}(- e^{\beta (\Delta - \mu)}) \right], \\
U_{2}^{(2)} 
&=& \frac{1}{m_{2}} D_2(\mu) (\Delta - \mu) \left( \frac{k_BT}{\Delta - \mu} \right)^{3/2} \nonumber \\
&& \times \left[
- \frac{35\sqrt{\pi}}{8} (k_BT)^{2}  \text{Li}_{\frac{7}{2}}(- e^{\beta (\Delta - \mu)}) 
+ \frac{5\sqrt{\pi}}{2}  k_BT (\Delta - \mu)  \text{Li}_{\frac{5}{2}}(- e^{\beta (\Delta - \mu)}) 
- \frac{\sqrt{\pi}}{2} (\Delta - \mu)^2  \text{Li}_{\frac{3}{2}}(- e^{\beta (\Delta - \mu)})  \right].
\end{eqnarray}
Note that $en_l = 2 m_l J_{1}^{(l)}$.
We can also evaluate $J^{(l)}_{i}$ and $U^{(l)}_{i}$ up to the leading order of temperature using the Sommerfeld expansion. For the electron band,
\begin{eqnarray}
J^{(1)}_{i} 
&\simeq&
\begin{cases}
\frac{2e}{3m_{1}} D_1(\mu) \mu  & \text{for}~i = 1, \\
\frac{\pi^2e}{3m_{1}} D_1(\mu)(k_BT)^{2} & \text{for}~i = 2, 
\end{cases} \label{J_el_low} \\
U^{(1)}_{i} 
&\simeq&
\begin{cases}
\frac{\pi^2}{3m_{1}} D_1(\mu) (k_BT)^{2} & \text{for}~i = 1  ,\\
\frac{2\pi^2}{9m_{1}} D_1(\mu) \mu (k_BT)^{2} & \text{for}~i = 2 ,\\
\end{cases}  \label{U_el_low}
\end{eqnarray}

and for the hole band,
\begin{eqnarray}
J^{(2)}_{i} 
&\simeq& 
\begin{cases}
\frac{2e}{3m_{2}} D_2(\mu) (\Delta - \mu)  & \text{for}~i = 1, \\
-\frac{\pi^2e}{3m_{2}} D_2(\mu)(k_BT)^{2} & \text{for}~i = 2,
\end{cases} \label{J_ho_low} \\
U^{(2)}_{i} 
&\simeq &
\begin{cases}
-\frac{\pi^2}{3m_{2}} D_2(\mu) (k_BT)^{2} & \text{for}~i = 1 , \\
\frac{2\pi^2}{9m_{2}} D_2(\mu) (\Delta - \mu) (k_BT)^{2} & \text{for}~i = 2. \\
\end{cases} \label{U_ho_low}
\end{eqnarray}
\subsection{Impurity scattering}
$P^{(11)}_{\text{imp},ij}$ is given by
\begin{eqnarray}
P^{(11)}_{\text{imp},ij}
&=& \frac{1}{2k_B TV} \sum_{\bm{k}_1,\bm{k}_3}
S_{\text{imp}}^{(l)}(\bm{k}_1,\bm{k}_3)  f_{0}(\varepsilon_{1,\bm{k}_1}) (1- f_{0}(\varepsilon_{1,\bm{k}_3})) \left( \frac{\hbar}{m_{1}} \right)^2   \nonumber \\ 
&&\times \left[
    k_{1;x}(\varepsilon_{1,\bm{k}_1} - \mu)^{i-1} - k_{3;x}(\varepsilon_{1,\bm{k}_3} - \mu)^{i-1}
 \right]
 \left[
    k_{1;x}(\varepsilon_{1,\bm{k}_3} - \mu)^{j-1} - k_{3;x}(\varepsilon_{1,\bm{k}_3} - \mu)^{j-1}
 \right] \nonumber \\
&=& \frac{1}{6k_B TV} \sum_{\bm{k}_1,\bm{k}_3}
S_0 \delta(\varepsilon_{1,\bm{k}_1} - \varepsilon_{1,\bm{k}_3}) f_{0}(\varepsilon_{1,\bm{k}_1}) (1- f_{0}(\varepsilon_{1,\bm{k}_1})) \left( \frac{\hbar}{m_{1}} \right)^2 (\bm{k}_{1} - \bm{k}_{3})^2(\varepsilon_{1,\bm{k}_1} - \mu)^{i + j - 2} \nonumber \\
&=& \frac{1}{6k_B T} \sum_{\bm{k}_1}
S_0 D(\varepsilon_{1,\bm{k}_1}) f_{0}(\varepsilon_{1,\bm{k}_1}) (1- f_{0}(\varepsilon_{1,\bm{k}_1})) \left( \frac{\hbar}{m_{1}} \right)^2 2\bm{k}_{1}^2 (\varepsilon_{1,\bm{k}_1} - \mu)^{i + j - 2} \nonumber \\
&=& \frac{2VS_0(D_1(\mu))^2\mu}{3m_{1}} \left( \frac{k_BT}{\mu} \right)^{2} \nonumber \\
&& \times
\begin{cases}
- 2\text{Li}_{2}(-e^{\beta \mu}) & \text{for}~ i = j = 1, \\
 - 6k_BT \text{Li}_{3}(-e^{\beta \mu}) + 2 \mu  \text{Li}_{2}(-e^{\beta \mu})  & \text{for}~ i = 1,j = 2 \text{~or~} i = 2,j = 1,\\
- 24(k_BT)^2 \text{Li}_{4}(-e^{\beta \mu}) + 12k_BT \mu\text{Li}_{3}(-e^{\beta \mu}) - 2\mu^2  \text{Li}_{2}(-e^{\beta \mu})   & \text{for}~ i = j = 2,\\
\end{cases} \\
&\simeq& \frac{2VS_0(D_1(\mu))^2\mu}{3m_{1}} \times
\begin{cases}
1 & \text{for}~ i = j = 1, \\
\frac{2\pi^2}{3\mu}(k_B T)^2   & \text{for}~ i = 1,j = 2 \text{~or~} i = 2,j = 1,\\
\frac{\pi^2}{3}(k_B T)^2   & \text{for}~ i = j = 2,\\
\end{cases}
\end{eqnarray}
and 
$P^{(22)}_{\text{imp},ij}$ can be calculated in this way,
\begin{eqnarray}
P^{(22)}_{\text{imp},ij}  
&=& \frac{2VS_0(D_2(\mu))^2(\Delta - \mu)}{3m_{2}} \left( \frac{k_BT}{\Delta - \mu} \right)^{2} \nonumber \\
&& \times
\begin{cases}
- 2\text{Li}_{2}(-e^{\beta(\Delta - \mu)}) & \text{for}~ i = j = 1, \\
 6k_BT \text{Li}_{3}(-e^{\beta(\Delta - \mu)}) - 2 (\Delta - \mu)  \text{Li}_{2}(-e^{\beta(\Delta - \mu)})  & \text{for}~ i = 1,j = 2 \text{~or~} i = 2,j = 1,\\
\begin{split}
  & - 24(k_BT)^2 \text{Li}_{4}(-e^{\beta(\Delta - \mu)}) + 12k_BT (\Delta - \mu)\text{Li}_{3}(-e^{\beta(\Delta - \mu)})   \\
  & \qquad - 2(\Delta - \mu)^2  \text{Li}_{2}(-e^{\beta(\Delta - \mu)})   
\end{split} & \text{for}~ i = j = 2,\\
\end{cases} \\
&\simeq& \frac{2VS_0(D_2(\mu))^2(\Delta - \mu)}{3m_{2}} \times
\begin{cases}
1 & \text{for}~i = j = 1, \\
-\frac{2\pi^2}{3(\Delta - \mu)}(k_B T)^2   & \text{for}~i = 1,j = 2 \text{~or~} i = 2,j = 1,\\
\frac{\pi^2}{3}(k_B T)^2   & \text{for}~i = j = 2.\\
\end{cases}
\end{eqnarray}
We note that the relaxation times $\tau^{(1)}_{\text{imp}}$ and $\tau^{(2)}_{\text{imp}}$ at $ T = 0$ are related to $S_0$ as
\begin{eqnarray}
V S_0 D_1(\varepsilon_{\text{F}}) \varepsilon_{\text{F}} &=& \frac{\varepsilon_{\text{F}}}{\tau^{(1)}_{\text{imp}}}, \\
V S_0 D_2(\varepsilon_{\text{F}}) (\Delta - \varepsilon_{\text{F}})  &=& \frac{\Delta - \varepsilon_{\text{F}}}{\tau^{(2)}_{\text{imp}}}.
\end{eqnarray}

\subsection{Electron-hole scattering}
The scattering matrices of the electron-hole scattering are given by eqs.~(B2) $\sim$ (B4) in Appendix B.
Due to the momentum conservation of the electron-hole scattering, we have relations as follows: 
\begin{eqnarray}
m_{1}^2P_{\text{e-h},11}^{(11)} = m_{2}^2P_{\text{e-h},11}^{(22)} &=& m_{1}m_{2}P_{\text{e-h},11}^{(12)} = m_{1}m_{2}P_{\text{e-h},11}^{(21)}, \label{constraint_P_11} \\
m_{1}P_{\text{e-h},21}^{(11)} &=& m_{2}P_{\text{e-h},21}^{(12)}, \label{constraint_P_21_1}  \\
m_{1}P_{\text{e-h},21}^{(21)} &=& m_{2}P_{\text{e-h},21}^{(22)}. \label{constraint_P_21_2} 
\end{eqnarray}

$P^{(11)}_{\text{e-h},ij}$ and $P^{(12)}_{\text{e-h},ij}$ are transformed as 
\begin{eqnarray}
P^{(11)}_{\text{e-h},ij} 
&=& \frac{1}{3k_B TV}\sum_{\bm{k}_1,\bm{k}_2,\bm{k}_3,\bm{k}_4} S_{\text{e-h}}(\bm{k}_1,\bm{k}_2;\bm{k}_3,\bm{k}_4) 
f_{0}(\varepsilon_{1,\bm{k}_1}) f_{0}(\varepsilon_{2,\bm{k}_2}) (1- f_{0}(\varepsilon_{1,\bm{k}_3})) (1- f_{0}(\varepsilon_{2,\bm{k}_4})) \nonumber \\ 
&& \times \Bigg[  \bm{v}^{(1)}_{\bm{k}_{1}} \cdot \bm{v}^{(1)}_{\bm{k}_1} (\varepsilon_{1,\bm{k}_1} - \mu)^{i-1}(\varepsilon_{1,\bm{k}_1} - \mu)^{j-1} - \bm{v}^{(1)}_{\bm{k}_{1}} \cdot \bm{v}^{(1)}_{\bm{k}_{3}} (\varepsilon_{1,\bm{k}_1} - \mu)^{i-1}(\varepsilon_{1,\bm{k}_{3}} - \mu)^{j-1} \nonumber \\
&& - \bm{v}^{(1)}_{\bm{k}_{3}} \cdot \bm{v}^{(1)}_{\bm{k}_1} (\varepsilon_{1,\bm{k}_{3}} - \mu)^{i-1}(\varepsilon_{1,\bm{k}_1} - \mu)^{j-1} + \bm{v}^{(1)}_{\bm{k}_{3}} \cdot \bm{v}^{(1)}_{\bm{k}_{3}} (\varepsilon_{{1},\bm{k}_{3}} - \mu)^{i-1}(\varepsilon_{{1},\bm{k}_{3}} - \mu)^{j-1} \Bigg] \nonumber \\
&=& \frac{2}{3k_B TV}\sum_{\bm{k}_1,\bm{k}_2,\bm{k}_3,\bm{k}_4} S_{\text{e-h}}(\bm{k}_1,\bm{k}_2;\bm{k}_3,\bm{k}_4) 
f_{0}(\varepsilon_{1,\bm{k}_1}) f_{0}(\varepsilon_{2,\bm{k}_2}) (1- f_{0}(\varepsilon_{1,\bm{k}_3})) (1- f_{0}(\varepsilon_{2,\bm{k}_4})) \nonumber \\ 
&& \times \Bigg[  \bm{v}^{(1)}_{\bm{k}_{1}} \cdot \bm{v}^{(1)}_{\bm{k}_1} (\varepsilon_{1,\bm{k}_1} - \mu)^{i + j - 2} - \bm{v}^{(1)}_{\bm{k}_{1}} \cdot \bm{v}^{(1)}_{\bm{k}_{3}} (\varepsilon_{1,\bm{k}_1} - \mu)^{i-1}(\varepsilon_{1,\bm{k}_{3}} - \mu)^{j-1}  \Bigg],
\end{eqnarray}
and 
\begin{eqnarray}
P^{(12)}_{\text{e-h},ij} 
&=& \frac{2}{3k_B TV}\sum_{\bm{k}_1,\bm{k}_2,\bm{k}_3,\bm{k}_4} S_{\text{e-h}}(\bm{k}_1,\bm{k}_2;\bm{k}_3,\bm{k}_4) 
f_{0}(\varepsilon_{1,\bm{k}_1}) f_{0}(\varepsilon_{2,\bm{k}_2}) (1- f_{0}(\varepsilon_{1,\bm{k}_3})) (1- f_{0}(\varepsilon_{2,\bm{k}_4})) \nonumber \\ 
&& \times \Bigg[  \bm{v}^{(1)}_{\bm{k}_{1}} \cdot \bm{v}^{(2)}_{\bm{k}_2} (\varepsilon_{1,\bm{k}_1} - \mu)^{i-1}(\varepsilon_{2,\bm{k}_2} - \mu)^{j-1} - \bm{v}^{(1)}_{\bm{k}_{1}} \cdot \bm{v}^{(2)}_{\bm{k}_{4}} (\varepsilon_{1,\bm{k}_1} - \mu)^{i-1}(\varepsilon_{2,\bm{k}_{4}} - \mu)^{j-1}  \Bigg],
\end{eqnarray}
where we have used the isotropy of the system and the relation
\begin{equation}
f_{0}(\varepsilon_{1,\bm{k}_1})f_{0}(\varepsilon_{2,\bm{k}_2})( 1 - f_{0}(\varepsilon_{1,\bm{k}_3}))(1 - f_{0}(\varepsilon_{2,\bm{k}_4})) = ( 1 - f_{0}(\varepsilon_{1,\bm{k}_1}))(1 - f_{0}(\varepsilon_{2,\bm{k}_2}))f_{0}(\varepsilon_{1,\bm{k}_3})f_{0}(\varepsilon_{2,\bm{k}_4}),
\end{equation}
which holds under the condition of energy conservation: $\varepsilon_{1,\bm{k}_1} + \varepsilon_{2,\bm{k}_2} = \varepsilon_{1,\bm{k}_3} + \varepsilon_{2,\bm{k}_4}$.

The momentum integral of $P^{(11)}_{\text{e-h},ij}$ is rather complicated, but angular integrals can be carried out analytically. First, $P^{(11)}_{\text{e-h},ij}$ can be rewritten as

\begin{eqnarray}
P^{(11)}_{\text{e-h},ij} 
&=& \frac{2}{3k_BT} \frac{2\pi}{\hbar}  \frac{\hbar^2}{m_{1}^2} \left( \frac{e^2}{\varepsilon_{0}} \right)^2 \frac{(2\pi)^3}{(4\pi)^4} \int \cdots \int d\varepsilon_{1} d\varepsilon_{2}d\varepsilon_{3} d\varepsilon_{4} D_{1}(\varepsilon_{1})D_{2}(\varepsilon_{2})D_{1}(\varepsilon_{3})D_{2}(\varepsilon_{4}) d\Omega_{\bm{k}_1}d\Omega_{\bm{k}_2}d\Omega_{\bm{k}_3}d\Omega_{\bm{k}_4} \int d^{3} \bm{q} \nonumber \\
&& \times \delta(\varepsilon_{1} + \varepsilon_{2} - \varepsilon_3 - \varepsilon_4)  f_{0}(\varepsilon_{1}) f_{0}(\varepsilon_{2}) (1- f_{0}(\varepsilon_{3})) (1- f_{0}(\varepsilon_{4})) \frac{1}{(q^2 + \alpha^2)^2} \nonumber \\
&& \times \delta(\bm{k}_1 - \bm{k}_3 - \bm{q}) \delta(\bm{k}_2 - \bm{k}_4 + \bm{q})\left[ k_1^2(\varepsilon_{1} - \mu)^{i + j -2} - \frac{k_1^2 + k_3^2 - q^2}{2} (\varepsilon_{1} - \mu)^{i - 1}(\varepsilon_3 - \mu)^{j - 1} \right].
\end{eqnarray}
where $k_{1(3)} = \sqrt{2m_{1}\varepsilon_{1(3)}/\hbar^2}$.
Choosing the direction of $\bm{q}$ as $z$ axis, the angular integral can be calculated~\cite{Maldague1979} as 
\begin{eqnarray}
&&\int d \Omega_{\bm{k}_2} \int d \Omega_{\bm{k}_4} \delta(\bm{k}_2 - \bm{k}_4 + \bm{q}) \nonumber \\
&=& \int d\Omega_{\bm{k}_2} \int d\bm{k}'_4 \frac{1}{k_4^2}\delta(k'_4 - k_4) \delta(\bm{k}_2 - \bm{k}'_4 + \bm{q}) \nonumber \\
&=& \int d\Omega_{\bm{k}_2} \frac{1}{k_4^2}\delta(|\bm{k}_2 + \bm{q}| - k_4)
\nonumber \\
&=& 2\pi \int_{0}^{\pi} d\theta \sin \theta \frac{1}{k_4^2} \delta \left( \sqrt{k_2^2 + q^2 + 2 k_2 q \cos \theta} - k_4 \right) =\frac{2\pi}{k_2 k_4 q} \Theta(k_2 + k_4 - q) \Theta(q - |k_2 - k_4|),
\end{eqnarray}
where $k_{2(4)} = \sqrt{2m_{2}(\Delta - \varepsilon_{2(4)})/\hbar^2}$. Note that we do not approximate $k_{1}$ and $k_{3}$ ($k_{2}$ and $k_{4}$) as $k_{\text{F},1}$ ($k_{\text{F},2}$) at this stage because the energy dependences of the wavenumbers are important when considering $L_{12}$, namely thermoelectric responses.
Then, $P^{(11)}_{\text{e-h},ij}$ becomes 
\begin{eqnarray}
P^{(11)}_{\text{e-h},ij} 
&=&  \frac{2P_{0,\text{e-h}}\alpha^3}{k_BT m_{1}^2} \int_{0}^{\infty} d\varepsilon_{1} \int_{-\infty}^{\Delta}d\varepsilon_{2} \int_{0}^{\infty} d\varepsilon_3 \int_{-\infty}^{\Delta} d\varepsilon_4 \int_{0}^{\infty} dq \delta(\varepsilon_{1} + \varepsilon_{2} - \varepsilon_{3} - \varepsilon_{4})  f_{0}(\varepsilon_{1}) f_{0}(\varepsilon_{2}) (1- f_{0}(\varepsilon_{3})) (1- f_{0}(\varepsilon_{4})) \nonumber \\
&&\times \frac{1}{(q^2 + \alpha^2)^2} \Theta(k_1 + k_3 - q) \Theta(q - |k_1 - k_3|)\Theta(k_2 + k_4 - q) \Theta(q - |k_2 - k_4|) \nonumber \\
&&\times \left[ k_1^2(\varepsilon_{1} - \mu)^{i + j -2} - \frac{k_1^2 + k_3^2 - q^2}{2} (\varepsilon_{1} - \mu)^{i - 1}(\varepsilon_3 - \mu)^{j - 1} \right], \label{P^11_ij_ene}
\end{eqnarray}
with
\begin{equation}
P_{0,\text{e-h}} = \frac{1}{3\cdot 2^4 \pi^5} \frac{e^4}{\varepsilon_{0}^2 \hbar^7} \frac{m_{1}^2 m_{2}^2}{\alpha^3}.
\end{equation}
The angular integral of $P^{(22)}_{\text{e-h},ij}$ can be calculated in the same way. $P^{(22)}_{\text{e-h},ij}$ is obtained from eq.~(\ref{P^11_ij_ene}) by replacing $m_{1}$ with $m_{2}$, and interchanging $k_1~(k_3)$ and $k_2~(k_4)$. 
Let us turn to the calculation of $P^{(12)}_{\text{e-h},ij}$,
\begin{eqnarray}
P^{(12)}_{\text{e-h},ij} 
&=& - \frac{2}{3k_BT} \frac{2\pi}{\hbar}  \frac{\hbar^2}{m_{1}m_{2}} \left( \frac{e^2}{\varepsilon_{0}} \right)^2 \frac{(2\pi)^3}{(4\pi)^4} \int \cdots \int d\varepsilon_{1} d\varepsilon_{2}d\varepsilon_{3} d\varepsilon_{4} D_{1}(\varepsilon_{1})D_{2}(\varepsilon_{2})D_{1}(\varepsilon_{3})D_{2}(\varepsilon_{4}) d\Omega_{\bm{k}_1}d\Omega_{\bm{k}_2}d\Omega_{\bm{k}_3}d\Omega_{\bm{k}_4} \int d^{3} \bm{q} \nonumber \\
&& \times \delta(\varepsilon_{1} + \varepsilon_{2} - \varepsilon_3 - \varepsilon_4)  f_{0}(\varepsilon_{1}) f_{0}(\varepsilon_{2}) (1- f_{0}(\varepsilon_{3})) (1- f_{0}(\varepsilon_{4})) \frac{1}{(q^2 + \alpha^2)^2} \delta(\bm{k}_1 - \bm{k}_3 - \bm{q}) \delta(\bm{k}_2 - \bm{k}_4 + \bm{q})  \nonumber \\
&& \times \left[ k_1 k_2 \cos \vartheta_{12}(\varepsilon_{1} - \mu)^{i - 1}(\varepsilon_{2} - \mu)^{j - 1} - k_1 k_4 \cos \vartheta_{14} (\varepsilon_{1} - \mu)^{i - 1}(\varepsilon_4 - \mu)^{j - 1} \right],
\end{eqnarray}
where $\vartheta_{12}~(\vartheta_{14})$ is the angle between $\bm{k}_1$ and $\bm{k}_2~(\bm{k}_4)$. 
The angular integral is performed in the same way as $P^{(12)}_{\text{e-h},ij}$, 
\begin{eqnarray}
&&\int \cdots \int d\Omega_{\bm{k}_1}d\Omega_{\bm{k}_2}d\Omega_{\bm{k}_3}d\Omega_{\bm{k}_4} \delta(\bm{k}_1 - \bm{k}_3 - \bm{q}) \delta(\bm{k}_2 - \bm{k}_4 + \bm{q}) \cos \vartheta_{12} \nonumber \\
&=& \int \cdots \int d\Omega_{\bm{k}_1} d\Omega_{\bm{k}_2} d\bm{k}'_3 d\bm{k}'_4 \frac{1}{k_3^2k_4^2} \delta(k'_3 -k_3)\delta(k'_4 -k_4)  \delta(\bm{k}_1 - \bm{k}'_3 - \bm{q}) \delta(\bm{k}_2 - \bm{k}'_4 + \bm{q})  \cos \vartheta_{12}  \nonumber \\
&=& \frac{1}{k_3^2k_4^2}\int  d\Omega_{\bm{k}_1} \int d\Omega_{\bm{k}_2}  \delta(|\bm{k}_1 - \bm{q}| - k_3)\delta(|\bm{k}_2 + \bm{q}| - k_4) \cos \vartheta_{12}
\nonumber \\
&=& \frac{1}{k_3^2k_4^2} \int d\theta_1 d\varphi_1 \sin \theta_1 \int d\theta_2 d\varphi_2 \sin \theta_2 \delta \left( \sqrt{k_1^2 + q^2 - 2 k_1 q \cos \theta_1} - k_3 \right) \delta \left( \sqrt{k_2^2 + q^2 + 2 k_2 q \cos \theta_2} - k_4 \right) \nonumber \\
&& \times ( \cos \theta_1 \cos \theta_2 + \sin \theta_1 \sin \theta_2 \cos (\varphi_2 - \varphi_1)) \nonumber \\
&=& \frac{4\pi^2}{k_3^2k_4^2} \left( \int_{-1}^{1} dx_1 \delta \left( \sqrt{k_1^2 + q^2 - 2 k_1 q x_1} - k_3 \right) x_1 \right) \times \left( \int_{-1}^{1} dx_2  \delta \left( \sqrt{k_2^2 + q^2 + 2 k_2 q x_2} - k_4 \right) x_2 \right)\nonumber \\ 
&=& \frac{\pi^2}{k_1^2k_2^2k_3k_4q^4} (k_1^2 - k_3^2 + q^2) (k_4^2 - k_2^2 - q^2) \Theta(k_1 + k_3 - q) \Theta(q - |k_1 - k_3|)\Theta(k_2 + k_4 - q) \Theta(q - |k_2 - k_4|).
\end{eqnarray}
Therefore, we obtain 
\begin{eqnarray}
P^{(12)}_{\text{e-h},ij} 
&=&  \frac{P_{0,\text{e-h}}\alpha^3}{2k_BT m_{1}m_{2}} \int_{0}^{\infty} d\varepsilon_{1} \int_{-\infty}^{\Delta}d\varepsilon_{2} \int_{0}^{\infty} d\varepsilon_3 \int_{-\infty}^{\Delta} d\varepsilon_4 \int_{0}^{\infty} dq \delta(\varepsilon_{1} + \varepsilon_{2} - \varepsilon_{3} - \varepsilon_{4})  f_{0}(\varepsilon_{1}) f_{0}(\varepsilon_{2}) (1- f_{0}(\varepsilon_{3})) (1- f_{0}(\varepsilon_{4})) \nonumber \\
&&\times \frac{1}{q^2(q^2 + \alpha^2)^2} \Theta(k_1 + k_3 - q) \Theta(q - |k_1 - k_3|)\Theta(k_2 + k_4 - q) \Theta(q - |k_2 - k_4|) \nonumber \\
&&\times \Bigg[ (q^2 + k_1^2 - k_3^2)(q^2 + k_2^2 - k_4^2)(\varepsilon_{1} - \mu)^{i-1}(\varepsilon_{2} - \mu)^{j-1}  \nonumber \\
&& + (q^2 + k_1^2 - k_3^2)(q^2 - k_2^2 + k_4^2) (\varepsilon_{1} - \mu)^{i-1}(\varepsilon_{4} - \mu)^{j-1} \Bigg].  \label{P^12_ij_ene}
\end{eqnarray}

We numerically evaluate the energy integrals of eqs.~(\ref{P^11_ij_ene}) and (\ref{P^12_ij_ene}) for the results in the main text.

The energy integrals can be also carried out analytically up to the leading order of temperature~\cite{Ziman2001}. For example, assuming that the dominant contribution is near $\varepsilon_{\text{F}}$,
\begin{eqnarray}
&&\int_{-\infty}^{\infty} \cdots \int_{-\infty}^{\infty} d\varepsilon_{1} d\varepsilon_{2} d\varepsilon_3 d\varepsilon_4 \delta(\varepsilon_{1} + \varepsilon_{2} - \varepsilon_3 - \varepsilon_4) f_{0}(\varepsilon_{1}) f_{0}(\varepsilon_{2}) (1- f_{0}(\varepsilon_{3})) (1- f_{0}(\varepsilon_{4})) (\varepsilon_{1} - \mu)^{i-1}(\varepsilon_{3} - \mu)^{j-1} \nonumber \\
&=&\int_{-\infty}^{\infty} \cdots \int_{-\infty}^{\infty} d\varepsilon_{1} d\varepsilon_{2} d\varepsilon_3 d\varepsilon_4 d\mathcal{E} \delta(\varepsilon_{1}  - \varepsilon_3 - \mathcal{E}) \delta(\varepsilon_{2}  - \varepsilon_4 + \mathcal{E}) f_{0}(\varepsilon_{1}) f_{0}(\varepsilon_{2}) (1- f_{0}(\varepsilon_{3})) (1- f_{0}(\varepsilon_{4})) (\varepsilon_{1} - \mu)^{i-1}(\varepsilon_{3} - \mu)^{j-1} \nonumber \\
&=&\int_{-\infty}^{\infty} \cdots \int_{-\infty}^{\infty} d\varepsilon_{1} d\varepsilon_{2}  d\mathcal{E}  f_{0}(\varepsilon_{1}) f_{0}(\varepsilon_{2}) (1- f_{0}(\varepsilon_{1} - \mathcal{E})) (1- f_{0}(\varepsilon_{2} + \mathcal{E})) (\varepsilon_{1} - \mu)^{i-1}(\varepsilon_{1} - \mathcal{E} - \mu)^{j-1} .
\end{eqnarray}
Using the integral by part, we can carry out the $\varepsilon_{1}~(\varepsilon_{2})$ integral as,
\begin{eqnarray}
\int_{-\infty}^{\infty} d\varepsilon_{1} (\varepsilon_{1} - \mu)^{l}  f_{0}(\varepsilon_{1})(1- f_{0}(\varepsilon_{1} - \mathcal{E})) 
&=& \frac{1}{e^{\beta \mathcal{E}} - 1} \int_{-\infty}^{\infty} d\varepsilon_{1} (\varepsilon_{1} - \mu)^{l} \left(\frac{1}{e^{\beta(\varepsilon_{1} -\mathcal{E} - \mu)} + 1} - \frac{1}{e^{\beta(\varepsilon_{1}  - \mu)} + 1}\right) \nonumber \\
&=& \frac{1}{e^{\beta \mathcal{E}} - 1} \int_{-\infty}^{\infty} d\varepsilon_{1} \frac{(\varepsilon_{1} - \mu)^{l + 1}}{ l + 1} \beta  \left( - \frac{e^{\beta(\varepsilon_{1} -\mathcal{E} - \mu)}}{(e^{\beta(\varepsilon_{1} -\mathcal{E} - \mu)} + 1)^2} + \frac{e^{\beta(\varepsilon_{1} - \mu)}}{(e^{\beta(\varepsilon_{1}  - \mu)} + 1)^2}\right) \nonumber \\ 
&=& \frac{1}{e^{\beta \mathcal{E}} - 1} \int_{-\infty}^{\infty} d\varepsilon_{1} \frac{ (\varepsilon_{1} + \mathcal{E} - \mu)^{l + 1}- (\varepsilon_{1} - \mu)^{l + 1}}{ l + 1}  \left( - \frac{\partial f_0(\varepsilon_{1})}{\partial \varepsilon_{1}}\right) \nonumber \\ 
&=& \frac{1}{e^{\beta \mathcal{E}} - 1} 
\begin{cases}
\mathcal{E} & \text{for}~l = 0, \\
\frac{\mathcal{E}^2}{2} & \text{for}~l = 1, \\
\frac{\mathcal{E}}{3}\left( \pi^2 (k_BT)^2 + \mathcal{E}^2 \right) & \text{for}~l = 2.
\end{cases} \label{e1_integral}
\end{eqnarray}
Then, $P^{(11)}_{\text{e-h},11}$ becomes 
\begin{eqnarray}
P^{(11)}_{\text{e-h},11} 
&\simeq& \frac{P_{0,\text{e-h}}\alpha^3}{k_BT m_{1}^2} \int_{-\infty}^{\infty} d\mathcal{E}\int_{0}^{\infty} dq \frac{q^2}{(q^2 + \alpha^2)^2}\Theta(2k_{\text{F},1}  - q)\Theta(2k_{\text{F},2}  - q) \frac{\mathcal{E}^2}{(e^{\beta \mathcal{E}} - 1)(1 - e^{ - \beta \mathcal{E}})} \nonumber \\
&=& \frac{2\pi^2P_{0,\text{e-h}} }{3m_{1}^2} \alpha^2 \mathcal{I}_{2}(2q_{\text{F}}/\alpha) (k_BT)^2,
\end{eqnarray}
where we have put $k_1 = k_3 = k_{\text{F},1} $, $k_2 = k_4 = k_{\text{F},2} $ and $q_{\text{F}} = \min \{k_{\text{F},1} ,k_{\text{F},2}  \}$, and we have used 
\begin{equation}
\int_{-\infty}^{\infty} d\mathcal{E} \frac{\mathcal{E}^n}{(e^{\beta \mathcal{E}} - 1)(1 - e^{ - \beta \mathcal{E}})} = \begin{cases}
\frac{2\pi^2}{3} (k_BT)^3 & \text{for}~n = 2, \\
\frac{8\pi^4}{15} (k_BT)^5 & \text{for}~n = 4. \label{e_integral}
\end{cases} 
\end{equation}
$\mathcal{I}_n(x)$ is defined as 
\begin{equation}
\mathcal{I}_{n}(x) = \int_{0}^{x} dz \frac{z^n}{(z^2+1)^2} 
= \begin{cases}
\frac{1}{2}\left( \frac{x}{x^2 + 1} + \arctan x \right) & \text{for}~n = 0, \\
\frac{1}{2}\left(- \frac{x}{x^2 + 1} + \arctan x \right)  & \text{for}~n = 2.
\end{cases}
\end{equation}
The other energy integrals for $P^{(lk)}_{\text{e-h},11}$ and $P^{(lk)}_{\text{e-h},22}$ is performed in the same way. The results are shown later.

\begin{figure}[tbp]
\begin{center}
\rotatebox{0}{\includegraphics[angle=0,width=0.75\linewidth]{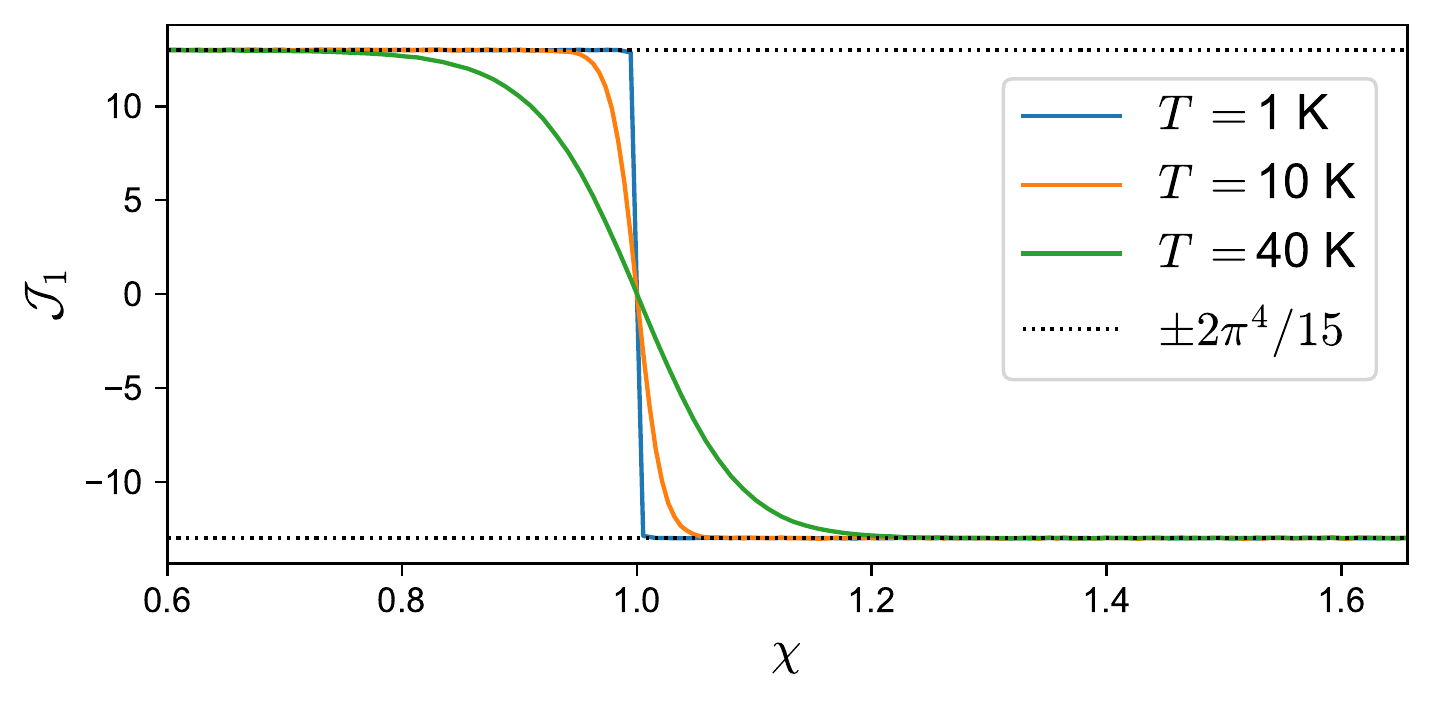}}
\caption{$\chi$ dependence of $\mathcal{J}_1(\chi,T)$ for three temperatures.}
\label{J_1_chi}
\end{center}
\end{figure}

In the case of $P^{(11)}_{\text{e-h},21}$, if we put $k_1 = k_3 = k_{\text{F},1} $ and $k_2 = k_4 = k_{\text{F},2} $, $P^{(11)}_{\text{e-h},21}$ vanishes because the energy integral in eq.~(\ref{P^12_ij_ene}) becomes an odd function of energy. Therefore, we have to take account of the $\varepsilon_{1}$ dependence of $k_1$, etc., as 
\begin{equation}
k_1 = \sqrt{\frac{2m_{1}}{\hbar^2}(\mu + (\varepsilon_1 - \mu))} \simeq k_{\text{F},1}  + \frac{k_{\text{F},1} }{2} \frac{\varepsilon_1 - \mu}{\mu}, \label{k_expanding}
\end{equation}
with $\mu = \hbar^2k_{\text{F},1}^2 /2m_{1}$.
We can evaluate a $q$ integral for an arbitrary function $f(q)$ and step functions as follows:
\begin{eqnarray}
&&\int_{0}^{\infty}  dq f(q) \Theta(k_1 + k_3 - q) \Theta(q - |k_1 - k_3|)\Theta(k_2 + k_4 - q) \Theta(q - |k_2 - k_4|) \nonumber \\
&=&\int_{0}^{2q_{\text{F}}}f(q) dq + \int_{0}^{\infty} dq f(q)( \Theta(k_1 + k_3 - q) \Theta(q - |k_1 - k_3|)\Theta(k_2 + k_4 - q) \Theta(q - |k_2 - k_4|) -\Theta(2k_{\text{F},1}  - q)\Theta(2k_{\text{F},2}  - q) ) \nonumber \\
&\simeq&\int_{0}^{2q_{\text{F}}}f(q) dq + \int_{\max\{|k_1 - k_3|,|k_2 - k_4|\}}^{\min\{k_1 + k_3,k_2 + k_4\}} f(q) dq  - \int_{0}^{2q_{\text{F}}} f(q) dq \nonumber \\
&\simeq& \int_{0}^{2q_{\text{F}}}f(q) dq + f(2q_{\text{F}})(\min\{k_1 + k_3,k_2 + k_4\} - 2q_{\text{F}})  -  f(0) \max\{|k_1 - k_3|,|k_2 - k_4|\}, \label{cutoff_q_integral}
\end{eqnarray}  
where $\min\{k_1 + k_3,k_2 + k_4\} - 2q_{\text{F}}$ and $\max\{|k_1 - k_3|,|k_2 - k_4|\}$ are in the order of $k_BT$.
Using eq.~(\ref{k_expanding}), the relations $\min\{x, y\} = (x + y - |x - y|)/2$, and $\max\{x, y\} = (x + y + |x - y|)/2$, we obtain
\begin{eqnarray}
&&\min\{k_1 + k_3,k_2 + k_4\} - 2q_{\text{F}}  \nonumber \\
&=& \frac{k_1 + k_3 + k_2 + k_4 - |k_1 + k_3 - k_2 - k_4|}{2} - \frac{2k_{\text{F},1} + 2k_{\text{F},2} - |2k_{\text{F},1} - 2k_{\text{F},2}|}{2} \nonumber \\
&=& \frac{k_1 + k_3 - 2k_{\text{F},1} + k_2 + k_4 - 2k_{\text{F},2}}{2} - \frac{|k_1 + k_3 - k_2 - k_4| - |2k_{\text{F},1} - 2k_{\text{F},2}|}{2}  \nonumber \\
&\simeq& k_{\text{F},1} \frac{\varepsilon_{1} + \varepsilon_3 - 2\mu }{4\mu} - k_{\text{F},2} \frac{\varepsilon_{2} + \varepsilon_4 - 2\mu }{4(\Delta - \mu)} \nonumber \\
&& - \left(\left|k_{\text{F},1} - k_{\text{F},2} + k_{\text{F},1} \frac{\varepsilon_{1} + \varepsilon_3 - 2\mu }{4\mu} +  k_{\text{F},2} \frac{\varepsilon_{2} + \varepsilon_4 - 2\mu }{4(\Delta - \mu)}\right|  - |k_{\text{F},1} - k_{\text{F},2}| \right), \label{min_decompose}
\end{eqnarray}
and 
\begin{eqnarray}
&&\max\{|k_1 - k_3|,|k_2 - k_4|\}  \nonumber \\
&\simeq& \max\left\{ \frac{k_{\text{F},1}}{2\mu}|\varepsilon_{1} - \varepsilon_3|,\frac{k_{\text{F},2}}{2(\Delta - \mu)}|\varepsilon_{2} - \varepsilon_4|\right\} \nonumber \\
&=& |\varepsilon_{1} - \varepsilon_3| \max\left\{ \frac{k_{\text{F},1}}{2\mu},\frac{k_{\text{F},2}}{2(\Delta - \mu)}\right\},
\end{eqnarray}
where the last equality follows from energy conservation. With these formulae, the $q$ integral in eq.~(\ref{P^11_ij_ene}) can be carried out as
\begin{eqnarray}
P^{(11)}_{\text{e-h},12} 
&\simeq& \frac{2P_{0,\text{e-h}}}{k_BT m_{1}^2} \alpha^{3}  \int \cdots \int d\varepsilon_{1} d\varepsilon_{2} d\varepsilon_3 d\varepsilon_4 \delta(\varepsilon_{1} + \varepsilon_{2} - \varepsilon_{3} - \varepsilon_{4})  f_{0}(\varepsilon_{1}) f_{0}(\varepsilon_{2}) (1- f_{0}(\varepsilon_{3})) (1- f_{0}(\varepsilon_{4})) \nonumber \\
&&\times \Biggl[ \alpha^{-3} \mathcal{I}_{0}(2q_{\text{F}}/\alpha) \left( k_1^2(\varepsilon_{1} - \mu) - \frac{k_1^2 + k_3^2 }{2} (\varepsilon_3 - \mu)\right) + \frac{\alpha^{-1}}{2} \mathcal{I}_{2}(2q_{\text{F}}/\alpha) (\varepsilon_3 - \mu) \nonumber \\
&&+ \frac{1}{((2q_{\text{F}})^2 + \alpha^2)^2} \left( k_1^2(\varepsilon_{1} - \mu) - \frac{k_1^2 + k_3^2 - (2q_{\text{F}})^2}{2} (\varepsilon_3 - \mu)  \right) (\min\{k_1 + k_3,k_2 + k_4\} - 2q_{\text{F}}) \nonumber \\
&&- \frac{1}{\alpha^4} \left( k_1^2(\varepsilon_{1} - \mu) - \frac{k_1^2 + k_3^2}{2}(\varepsilon_3 - \mu)  \right) \max\{|k_1 - k_3|,|k_2 - k_4|\}\Biggr] \nonumber \\
&\simeq& \frac{2P_{0,\text{e-h}}}{k_BT m_{1}^2} \alpha^{3}  \int \cdots \int d\varepsilon_{1} d\varepsilon_{2} d\varepsilon_3 d\varepsilon_4 \delta(\varepsilon_{1} + \varepsilon_{2} - \varepsilon_{3} - \varepsilon_{4})  f_{0}(\varepsilon_{1}) f_{0}(\varepsilon_{2}) (1- f_{0}(\varepsilon_{3})) (1- f_{0}(\varepsilon_{4})) \nonumber \\
&&\times \Biggl[ \alpha^{-3} \mathcal{I}_{0}(2q_{\text{F}}/\alpha)  \frac{k_{\text{F},1}^2}{\mu}  \left( (\varepsilon_{1} - \mu)^2 - \frac{(\varepsilon_{1} + \varepsilon_3 - 2\mu)}{2} (\varepsilon_3 - \mu)\right) \nonumber \\
&& +\frac{1}{((2q_{\text{F}})^2 + \alpha^2)^2} \left( k_{\text{F},1}^2 (\varepsilon_{1} - \varepsilon_3) + 2q_{\text{F}}^2(\varepsilon_3 - \mu) \right)(\min\{k_1 + k_3,k_2 + k_4\} - 2q_{\text{F}}) \Biggr],
\end{eqnarray}
where we keep only the lowest even order of energy. Note that the contribution from $q \sim 0$ vanishes since $(\varepsilon_3 - \mu)\max\{|k_1 - k_3|,|k_2 - k_4|\} \propto (\varepsilon_3 - \mu)|\mathcal{E}|$ is a odd function of energy. Now, we can carry out the energy integrals using eqs.~(\ref{e1_integral}), (\ref{e_integral}), and functions $\mathcal{J}_1(\chi,T)$ and $\mathcal{J}_2(\chi,T)$ defined as
\begin{eqnarray}
\mathcal{J}_1(\chi,T) =  \frac{\mu}{k_BT} \int_{-\infty}^{\infty} dx \int_{-\infty}^{\infty} d\eta_1 \int_{-\infty}^{\infty} d\eta_2 &&\frac{\eta_1}{(e^{\eta_1 + x/2} + 1)(e^{-\eta_1 + x/2} + 1)(e^{\eta_2 - x/2} + 1)(e^{-\eta_2 - x/2} + 1)} \nonumber \\
&& \times \left|1 - \chi + \frac{k_BT}{2\mu} \eta_1 + \frac{\chi k_BT}{2(\Delta - \mu)}\eta_2 \right|, \\
\mathcal{J}_2(\chi,T) = \frac{\Delta - \mu}{k_BT}\int_{-\infty}^{\infty} dx \int_{-\infty}^{\infty} d\eta_1 \int_{-\infty}^{\infty} d\eta_2 &&\frac{\eta_2}{(e^{\eta_1 + x/2} + 1)(e^{-\eta_1 + x/2} + 1)(e^{\eta_2 - x/2} + 1)(e^{-\eta_2 - x/2} + 1)} \nonumber \\
&& \times \left|\chi^{-1} - 1 + \frac{\chi^{-1}k_BT}{2\mu} \eta_1 + \frac{k_BT}{2(\Delta - \mu)}\eta_2 \right|,
\end{eqnarray}
which correspond to the energy integral for the second term of eq.~(\ref{min_decompose}).
$\chi$ dependences of $\mathcal{J}_1(1,T)$ for $T = 1$, $10$, and $40~\text{K}$ are shown in Fig.~\ref{J_1_chi} setting $\Delta = 0.2~\text{eV}$ and $m_{2} = 3m_{1} = 3m_{\text{e}}$. We can analytically evaluate the limiting values as 
\begin{eqnarray}
\mathcal{J}_1(\chi \to 0 ,T) = \mathcal{J}_2(\chi \to 0 ,T) = - \mathcal{J}_1(\chi \to \infty,T) = - \mathcal{J}_2(\chi \to \infty,T) = \frac{2\pi^4}{15},
\end{eqnarray}
and we have
\begin{equation}
\mathcal{J}_1(1,T) = \mathcal{J}_2(1,T) = 0.
\end{equation}

The matrix elements $P^{(11)}_{\text{e-h},ij}$ up to the leading order of temperature are as follows:
\begin{eqnarray}
P^{(11)}_{\text{e-h},11} &\simeq& \frac{2\pi^2P_{0,\text{e-h}} }{3m_{1}^2} \alpha^2 \mathcal{I}_{2}(2q_{\text{F}}/\alpha) (k_BT)^2 , \\
P^{(11)}_{\text{e-h},21} = P^{(11)}_{\text{e-h},12} &\simeq& \frac{4\pi^4 P_{0,\text{e-h}} }{15 m_{1}^2} \frac{k_{\text{F,1}}^2}{\mu} \mathcal{I}_{0}(2q_{\text{F}}/\alpha) (k_BT)^4 \nonumber \\
&& + \frac{P_{0,\text{e-h}}}{m_{1}^2} \frac{\alpha^3(2q_{\text{F}})^2 }{\left((2q_{\text{F}})^2 + \alpha^2\right)^2} \frac{k_{\text{F},1}}{\mu}  \left( \frac{2\pi^4}{15} - \mathcal{J}_1(\chi,T) \right)  (k_BT)^4, \\
P^{(11)}_{\text{e-h},22} &\simeq& \frac{P_{0,\text{e-h}} }{m_{1}^2} 
\left[
  \frac{8\pi^4}{15}  k_{\text{F},1}^2 \mathcal{I}_{0}(2q_{\text{F}}/\alpha) + \frac{2\pi^4}{15}\alpha^2\mathcal{I}_{2}(2q_{\text{F}}/\alpha)
\right] (k_BT)^4 ,
\end{eqnarray}

The results for $P^{(22)}_{\text{e-h},ij}$ and $P^{(12)}_{\text{e-h},ij}$ are similar to those of $P^{(11)}_{\text{e-h},ij}$,
\begin{eqnarray}
P^{(22)}_{\text{e-h},11} &\simeq& \frac{2\pi^2P_{0,\text{e-h}} }{3m_{2}^2} \alpha^2 \mathcal{I}_{2}(2q_{\text{F}}/\alpha) (k_BT)^2 , \\
P^{(22)}_{\text{e-h},21} = P^{(22)}_{\text{e-h},12} &\simeq& - \frac{4\pi^4 P_{0,\text{e-h}} }{15 m_{2}^2} \frac{k_{\text{F,2}}^2}{\Delta - \mu} \mathcal{I}_{0}(2q_{\text{F}}/\alpha) (k_BT)^4 \nonumber \\
&& - \frac{P_{0,\text{e-h}}}{m_{2}^2} \frac{\alpha^3(2q_{\text{F}})^2 }{\left((2q_{\text{F}})^2 + \alpha^2\right)^2} \frac{k_{\text{F},2}}{\Delta - \mu}  \left(  \frac{2\pi^4}{15} + \mathcal{J}_2(\chi,T) \right)  (k_BT)^4, \\
P^{(22)}_{\text{e-h},22} &\simeq& \frac{P_{0,\text{e-h}} }{m_{2}^2} 
\left[
  \frac{8\pi^4}{15}  k_{\text{F},2}^2 \mathcal{I}_{0}(2q_{\text{F}}/\alpha) + \frac{2\pi^4}{15}\alpha^2\mathcal{I}_{2}(2q_{\text{F}}/\alpha)
\right] (k_BT)^4 ,
\end{eqnarray}
\begin{eqnarray}
P^{(12)}_{\text{e-h},11} &\simeq& \frac{2\pi^2P_{0,\text{e-h}} }{3m_{1}m_{2}} \alpha^2 \mathcal{I}_{2}(2q_{\text{F}}/\alpha) (k_BT)^2 , \\
P^{(12)}_{\text{e-h},22} & \simeq& 0, \\
P^{(12)}_{\text{e-h},21} &\simeq& \frac{4\pi^4 P_{0,\text{e-h}} }{15 m_{1}m_{2}} \frac{k_{\text{F,1}}^2}{\mu} \mathcal{I}_{0}(2q_{\text{F}}/\alpha) (k_BT)^4 \nonumber \\
&& + \frac{P_{0,\text{e-h}}}{ m_{1}m_{2}} \frac{\alpha^3(2q_{\text{F}})^2 }{\left((2q_{\text{F}})^2 + \alpha^2\right)^2} \frac{k_{\text{F},1}}{\mu}  \left( \frac{2\pi^4}{15} - \mathcal{J}_1(\chi,T) \right)  (k_BT)^4 \\
P^{(12)}_{\text{e-h},12} &\simeq& - \frac{4\pi^4 P_{0,\text{e-h}} }{15 m_{1}m_{2}} \frac{k_{\text{F,2}}^2}{\Delta - \mu} \mathcal{I}_{0}(2q_{\text{F}}/\alpha) (k_BT)^4 \nonumber \\
&& - \frac{P_{0,\text{e-h}}}{ m_{1}m_{2}} \frac{\alpha^3(2q_{\text{F}})^2 }{\left((2q_{\text{F}})^2 + \alpha^2\right)^2} \frac{k_{\text{F},2}}{\Delta - \mu}  \left(  \frac{2\pi^4}{15} + \mathcal{J}_2(\chi,T) \right)  (k_BT)^4.
\end{eqnarray} 

\subsection{Intraband scattering}
Scattering matrices of the intraband scattering are given by eq.~(\ref{scat_P_ee}).
Since we consider the model with quadratic dispersions, $P^{(ll)}_{\text{e-e},11},P^{(ll)}_{\text{e-e},12}$, and $P^{(ll)}_{\text{e-e},21}$ vanish due to the momentum conservation and the only non-vanishing term is $P^{(ll)}_{\text{e-e},22}$ within our framework. The angular integration of $P^{(11)}_{\text{e-e},22}$ is done in a similar way as the interband scattering, 
\begin{eqnarray}
P^{(11)}_{\text{e-e},22} 
&=&  \frac{P^{(1)}_{0,\text{e-e}}\alpha^3}{2k_BT m_{1}^2} \int_{0}^{\infty} \cdots \int_{0}^{\infty} d\varepsilon_{1} d\varepsilon_{2}  d\varepsilon_3 d\varepsilon_4 \int_{0}^{\infty} dq \delta(\varepsilon_{1} + \varepsilon_{2} - \varepsilon_{3} - \varepsilon_{4})  f_{0}(\varepsilon_{1}) f_{0}(\varepsilon_{2}) (1- f_{0}(\varepsilon_{3})) (1- f_{0}(\varepsilon_{4})) \nonumber \\
&&\times \frac{1}{q^2(q^2 + \alpha^2)^2} \Theta(k_1 + k_3 - q) \Theta(q - |k_1 - k_3|)\Theta(k_2 + k_4 - q) \Theta(q - |k_2 - k_4|) \nonumber \\
&&\times \Bigg[ 
4k_{1}^2q^2 (\varepsilon_{1} - \mu)^{2} -  2(k_{1}^2 + k_{3}^2 - q^2) q^2 (\varepsilon_{1} - \mu)(\varepsilon_{3} - \mu) - (q^2 + k_1^2 - k_3^2)(q^2 + k_2^2 - k_4^2)(\varepsilon_{1} - \mu)(\varepsilon_{2} - \mu)  \nonumber \\
&& - (q^2 + k_1^2 - k_3^2)(q^2 - k_2^2 + k_4^2) (\varepsilon_{1} - \mu)(\varepsilon_{4} - \mu)\Bigg], \label{P^l_22_ene}
\end{eqnarray}
where $k_i = \sqrt{2m_{1}\varepsilon_i/\hbar^2}$ and 
\begin{equation}
P^{(l)}_{0,\text{e-e}} = \frac{1}{3\cdot 2^4 \pi^5} \frac{e^4}{\varepsilon_{0}^2 \hbar^7} \frac{m_{l}^4}{\alpha^3}.
\end{equation} $P^{(22)}_{\text{e-e},22}$ is obtained from eq.~(\ref{P^l_22_ene}) by replacing $m_{1}$, $\int_{0}^{\infty} d\varepsilon_i$, and $k_i = \sqrt{2m_{1}\varepsilon_i/\hbar^2}$ with $m_{2}$, $\int_{-\infty}^{\Delta} d\varepsilon_i$ and $k_i = \sqrt{2m_{2}(\Delta - \varepsilon_i)/\hbar^2}$. The low-temperature expansion gives
\begin{equation}
P^{(ll)}_{\text{e-e},22}  \simeq  \frac{P^{(l)}_{0,\text{e-e}}}{m_l^2} 
\left[
  \frac{8\pi^4}{15}  k_{\text{F},l}^2 \mathcal{I}_{0}(2k_{\text{F},l}/\alpha) + \frac{2\pi^4}{15}\alpha^2\mathcal{I}_{2}(2k_{\text{F},l}/\alpha)
\right] (k_BT)^4.
\end{equation}
\begin{figure}[h]
\begin{center}
\rotatebox{0}{\includegraphics[angle=0,width=0.60\linewidth]{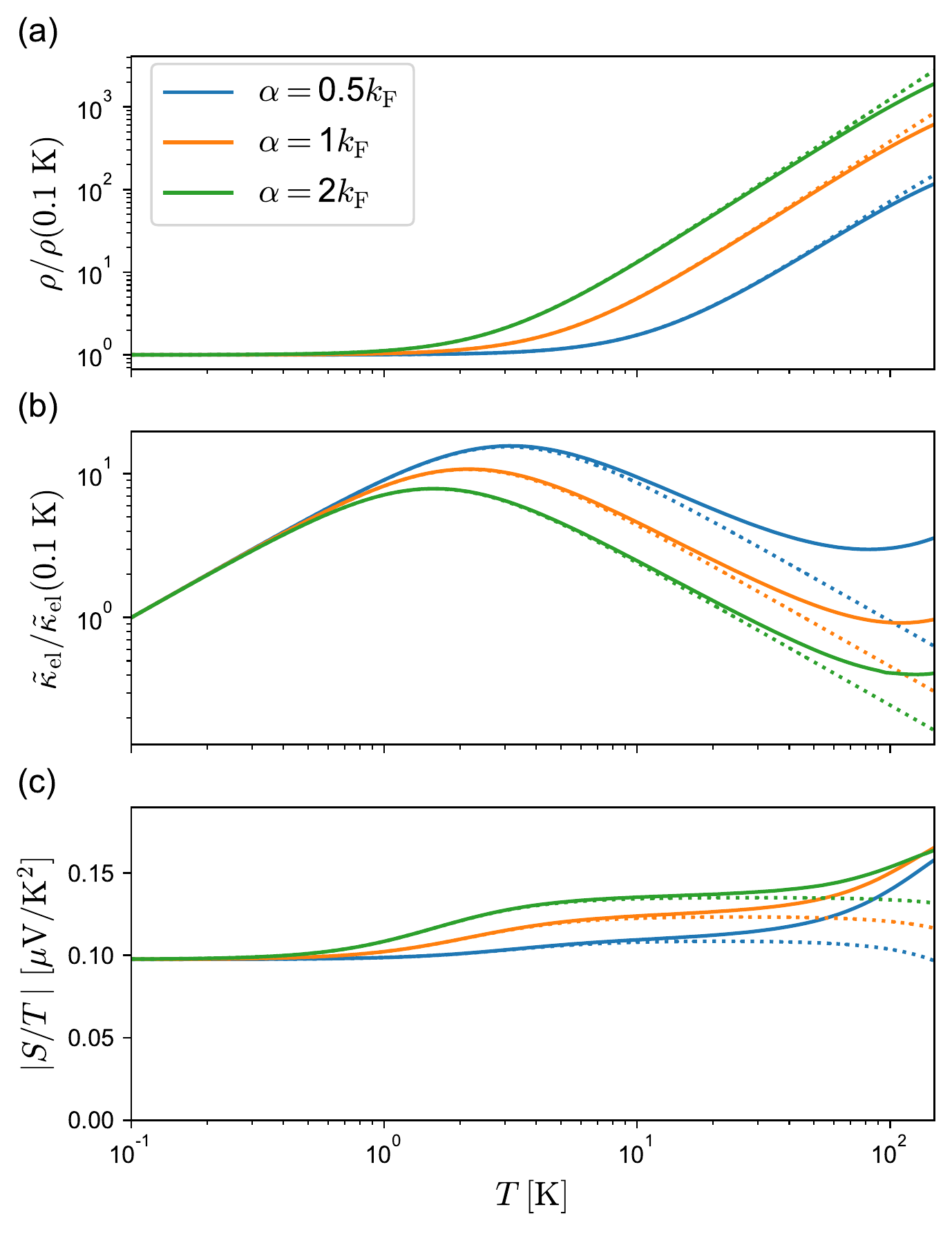}}
\caption{Temperature dependences of (a) electrical resistivity,  (b) thermal conductivity excluding the ambipolar contribution ($\tilde{\kappa}_{\text{el}}$) , and (c) $|S/T|$ for three screening lengths. Parameters are the same as in Fig.~3(a) in the main text. The solid lines show the results obtained via the numerical energy integral for $P_{ij}^{(lk)}$. The dotted lines show the results obtained via the low-temperature expansion for $P_{ij}^{(lk)}$.} 
\label{low_temp_expansion}
\end{center}
\end{figure}

\subsection{Comparison between numerical calculation and low-temperature expansion}
Now we have the analytical expressions of $P^{(lk)}_{ij}$ up to the leading order of temperature. Let us compare two ways of integral.
We show in Fig.~\ref{low_temp_expansion} the electrical resistivity, the thermal conductivity excluding the ambipolar contribution ($\tilde{\kappa}_{\text{el}} = 2({}^{t}\bm{U}_2 P_{22}^{-1}\bm{U}_2 )/T$), and $|S/T|$ for the compensated case using the numerical energy integral (solid lines) and the low-temperature expansion (dotted lines) for $P_{ij}^{(lk)}$. Parameters are the same as in Fig.~3(a) in the main text. While the electrical resistivity agrees well over the entire temperature range, the thermal conductivity deviates from the numerical calculations in the high temperature regime. We also observe that smaller $\alpha$ brings deviations in lower temperatures. This behavior is due to the sub-leading temperature dependence originating from the $q$ integral around $q \sim 0$ for $P^{(lk)}_{\text{e-h},22}$ and $P^{(lk)}_{\text{e-e},22}$. As seen in eq.~(\ref{cutoff_q_integral}), the sub-leading correction arising from the lower bound of the $q$ integral is $f(0)\min \{|k_1 - k_3|, |k_2 - k_4|\}$ for an arbitrary function $f(q)$. For the evaluation of leading order of $P^{(lk)}_{\text{e-h},22}$, we have a $q$ integral corresponding to $f(q) \propto (q^2 + \alpha^2)^{-2}$, which does not appear in $P^{(lk)}_{\text{e-h},11}$. Since $(q^2 + \alpha^2)^{-2} \to \alpha^{-4}~(q \to 0)$, the correction attributed to the integral around $q \sim 0$ can be large for small $\alpha$ and brings earlier deviations from expected temperature scaling for $\tilde{\kappa}_{\text{el}}$ compared to $\rho$.

\setcounter{equation}{0}
\setcounter{figure}{0}
\setcounter{table}{0}
\section{Relaxation time approximation (RTA)}
In this section, we discuss the relaxation time approximation (RTA) following Refs.~\cite{Lee2021, Nguyen2020}. We also discuss the behavior of the ambipolar contribution based on the obtained formulae.

\subsection{Derivation of transport coefficients}
The Boltzmann equation is given by
\begin{eqnarray}
\left(- e\bm{E} - (\varepsilon_{l,\bm{k}} - \mu)\left(-\frac{\nabla T}{T}\right) \right)\cdot \bm{v}^{(l)}_{\bm{k}} \left( -\frac{\partial f_{0}(\varepsilon_{l,\bm{k}})}{\partial \varepsilon_{l,\bm{k}}}\right) &=& - \frac{\delta f^{(l)}(\bm{k})}{\tau^{(l)}_{\text{imp}}} + \left(\frac{\partial f^{(l)}(\bm{k})}{\partial t} \right)_{\text{e-h}} + \left(\frac{\partial f^{(l)}(\bm{k})}{\partial t} \right)_{\text{e-e}}, \label{RTA_boltzmann}
\end{eqnarray}
with $\delta f^{(l)}(\bm{k}) = f^{(l)}(\bm{k}) - f_{0}(\varepsilon_{l,\bm{k}})$. For simplicity, we neglect the momentum dependence of $\tau^{(l)}_{\text{imp}}$.
The electron-hole scattering can be approximated in the RTA as 
\begin{eqnarray}
\left(\frac{\partial f^{(1)}(\bm{k})}{\partial t} \right)_{\text{e-h}} &=& - \frac{1}{\tau^{(12)}_{\text{e-h}}} \left[ \delta f^{(1)}(\bm{k}) + \bm{v}_{\text{d}}^{(2)} \cdot \hbar \bm{k} \left( - \frac{\partial  f_{0}(\varepsilon_{1,\bm{k}}) }{\partial \varepsilon_{1,\bm{k}}} \right) \right], \\
\left(\frac{\partial f^{(2)}(\bm{k})}{\partial t} \right)_{\text{e-h}} &=& -  \frac{1}{\tau^{(21)}_{\text{e-h}}} \left[ \delta f^{(2)}(\bm{k}) - \bm{v}_{\text{d}}^{(1)} \cdot \hbar \bm{k} \left( - \frac{\partial  f_{0}(\varepsilon_{2,\bm{k}}) }{\partial \varepsilon_{2,\bm{k}}} \right) \right],
\end{eqnarray}
and the intraband scattering can be approximated in the RTA as 
\begin{eqnarray}
\left(\frac{\partial f^{(1)}(\bm{k})}{\partial t} \right)_{\text{e-e}}  &=& -  \frac{1}{\tau^{(1)}_{\text{e-e}}} \left[ \delta f^{(1)}(\bm{k}) - \bm{v}_{\text{d}}^{(1)} \cdot \hbar \bm{k} \left( - \frac{\partial  f_{0}(\varepsilon_{1,\bm{k}}) }{\partial \varepsilon_{1,\bm{k}}} \right) \right] , \\
\left(\frac{\partial f^{(2)}(\bm{k})}{\partial t} \right)_{\text{e-e}}  &=& -  \frac{1}{\tau^{(2)}_{\text{e-e}}} \left[ \delta f^{(2)}(\bm{k}) + \bm{v}_{\text{d}}^{(2)} \cdot \hbar \bm{k} \left( - \frac{\partial  f_{0}(\varepsilon_{2,\bm{k}}) }{\partial \varepsilon_{2,\bm{k}}} \right) \right],
\end{eqnarray}
where the velocity $\bm{v}_{\text{d}}^{(l)}$, which we have to determine self-consistently, is given by
\begin{equation}
\bm{v}_{\text{d}}^{(l)} = \frac{2}{n_{l}V} \sum_{\bm{k}} \bm{v}^{(l)}_{\bm{k}}  f^{(l)}(\bm{k}) = \frac{2}{n_{l}V} \sum_{\bm{k}} \bm{v}^{(l)}_{\bm{k}}  \delta f^{(l)}(\bm{k}).
\end{equation}
The total momentum conservation requires a constraint on relaxation times,
\begin{equation}
\frac{m_{1}  n_{1}}{\tau^{(12)}_{\text{e-h}}} = \frac{m_{2}  n_{2}}{\tau^{(21)}_{\text{e-h}}}. \label{constraint_RTA}
\end{equation}
The temperature dependences of the relaxation times are as follows 
\begin{eqnarray}
\tau^{(1)}_{\text{imp}}, \tau^{(2)}_{\text{imp}} &\propto& T^{0} \\
\tau^{(12)}_{\text{e-h}}, \tau^{(21)}_{\text{e-h}} ,\tau^{(1)}_{\text{e-e}} ,\tau^{(2)}_{\text{e-e}} &\propto&{} T^{-2}
\end{eqnarray}
The electric current $\bm{J}$ is expressed by drift velocities as 
\begin{equation}
\bm{J} = \frac{2}{V} \sum_{l,\bm{k}} e \bm{v}^{(l)}_{\bm{k}} f^{(l)}(\bm{k}) = en_{1} \bm{v}_{\text{d}}^{(1)} + en_{2} \bm{v}_{\text{d}}^{(2)}.
\end{equation}
Note that, as discussed in Ref. \cite{Kukkonen1976}, the electric current is also expressed by the total momentum $\bm{P}$ and the relative momentum $\bm{p}_{\text{rel}}$ as
\begin{align}
\bm{J} =e \left( \frac{\bm{P}^{(1)}}{m_1} -  \frac{\bm{P}^{(2)}}{m_2}\right) =  e \left[ \frac{n_1 -n_2}{n_1m_1 + n_2m_2} \bm{P} + \left(\frac{1}{m_1} + \frac{1}{m_2}  \right) \bm{p}_{\text{rel}} \right],
\end{align}
where
\begin{align}
\bm{P}^{(l)} &= \frac{2}{V} \sum_{\bm{k}} \hbar \bm{k} f^{(l)}(\bm{k}), \\
\bm{P} &= \bm{P}^{(1)} + \bm{P}^{(2)}, \\
\bm{p}_{\text{rel}} &= \left( \frac{1}{n_1m_1} + \frac{1}{n_2m_2} \right)^{-1} \left( \frac{\bm{P}^{(1)}}{n_1m_1} - \frac{\bm{P}^{(2)}}{n_2m_2}  \right) = \left( \frac{1}{n_1m_1} + \frac{1}{n_2m_2} \right)^{-1} \left( \bm{v}_{\text{d}}^{(1)} + \bm{v}_{\text{d}}^{(2)} \right).
\end{align}
We see that the contribution from the total momentum to the electric current is proportional to $n_1 - n_2$.

By multiplying eq.~(\ref{RTA_boltzmann}) by $\bm{k}$ and integrating with respect to $\bm{k}$, we obtain the equation which determines $\bm{v}_{\text{d}}^{(l)}$ as
\begin{equation}
\left(
\begin{array}{cc}
\frac{1}{\tau^{(1)}_{\text{imp}}} + \frac{1}{\tau^{(12)}_{\text{e-h}}} & \frac{m_{1}n_{1}}{m_{2}n_{2}\tau^{(12)}_{\text{e-h}}} \\
\frac{m_{2}n_{2}}{m_{1}n_{1}\tau^{(21)}_{\text{e-h}}} & \frac{1}{\tau^{(2)}_{\text{imp}}} + \frac{1}{\tau^{(21)}_{\text{e-h}}}  \\
\end{array} \label{drift_v}
\right)
\left(
\begin{array}{c}
n_{1}m_{1}\bm{v}_{\text{d}}^{(1)}  \\
n_{2}m_{2}\bm{v}_{\text{d}}^{(2)}  \\
\end{array}
\right) =
\left(
\begin{array}{c}
en_{1}\bm{E} +  \braket{\varepsilon_{1,\bm{k}} - \mu}_1 n_{1} \left(-\frac{\nabla T}{T}\right) \\
en_{2}\bm{E} +  \braket{\varepsilon_{2,\bm{k}} - \mu}_2n_{2}\left(-\frac{\nabla T}{T}\right)   \\
\end{array}
\right),
\end{equation}
where $\braket{A_{\bm{k}}}_{l}$ denotes
\begin{equation}
\braket{A_{\bm{k}}}_{l} = \frac{2m_l}{3n_l} \frac{1}{V} \sum_{\bm{k}} (\bm{v}^{(l)}_{\bm{k}})^2 A_{\bm{k}} \left( -\frac{\partial f_{0}(\varepsilon_{l,\bm{k}})}{\partial \varepsilon_{l,\bm{k}}}\right),
\end{equation}
and satisfies $\braket{1}_{l} = 1$.
From eq.~(\ref{drift_v}), $\bm{v}_{\text{d}}^{(l)}$ is obtained. Therefore, $L^{(\text{RTA})}_{11}$ and $L^{(\text{RTA})}_{12} = L^{(\text{RTA})}_{21} $ are calculated as 
\begin{eqnarray}
L^{(\text{RTA})}_{11} &=& e^2 \left(\frac{1}{\tau^{(1)}_{\text{imp}}\tau^{(2)}_{\text{imp}}} + \frac{1}{\tau^{(12)}_{\text{e-h}}\tau^{(2)}_{\text{imp}}} + \frac{1}{\tau^{(21)}_{\text{e-h}}\tau^{(1)}_{\text{imp}}} \right)^{-1}
\left[  
 \frac{n_{1}}{m_{1}\tau^{(2)}_{\text{imp}}} + \frac{n_{2} - n_{1}}{m_{2}\tau^{(12)}_{\text{e-h}}} + \frac{n_{1} - n_{2}}{m_{1}\tau^{(21)}_{\text{e-h}}} + \frac{n_{2}}{m_{2}\tau^{(1)}_{\text{imp}}}
\right], \label{L_11_RTA}\\
L^{(\text{RTA})}_{12} &=& e \left(\frac{1}{\tau^{(1)}_{\text{imp}}\tau^{(2)}_{\text{imp}}} + \frac{1}{\tau^{(12)}_{\text{e-h}}\tau^{(2)}_{\text{imp}}} + \frac{1}{\tau^{(21)}_{\text{e-h}}\tau^{(1)}_{\text{imp}}} \right)^{-1} \nonumber \\
&&\times
\left[  
 \frac{n_{1}\braket{\varepsilon_{1,\bm{k}} - \mu}_1}{m_{1}\tau^{(2)}_{\text{imp}}} + \frac{(n_{2} - n_{1})\braket{\varepsilon_{2,\bm{k}} - \mu}_2}{m_{2}\tau^{(12)}_{\text{e-h}}} + \frac{(n_{1} - n_{2})\braket{\varepsilon_{1,\bm{k}} - \mu}_1}{m_{1}\tau^{(21)}_{\text{e-h}}} + \frac{n_{2}\braket{\varepsilon_{2,\bm{k}} - \mu}_2}{m_{2}\tau^{(1)}_{\text{imp}}} \right]. \label{L_12_RTA}
\end{eqnarray}
By setting $n_{1} = n_{2}$, we see that $S^{(\text{RTA})} = L^{(\text{RTA})}_{12}/TL^{(\text{RTA})}_{11}$ is independent of $\tau^{(12)}_{\text{e-h}}$ and $\tau^{(21)}_{\text{e-h}}$. 

By taking the limit of $\tau^{(12)}_{\text{e-h}}, \tau^{(21)}_{\text{e-h}} \ll \tau^{(1)}_{\text{imp}}, \tau^{(2)}_{\text{imp}}$ in eq. (\ref{L_11_RTA}), the saturated resistivity $\rho_{\text{sat}}$ is calculated as
\begin{eqnarray}
\rho_{\text{sat}} &=& \frac{1}{\frac{e^2(n_2 - n_1)^2}{m_1 n_1 /\tau_{\text{imp}}^{(1)} + m_2 n_2 /\tau_{\text{imp}}^{(2)}}} \propto \frac{1}{(n_2 - n_1)^2}.
\end{eqnarray}
$\rho_{\text{sat}}/\rho(T= 0)$ becomes
\begin{eqnarray}
\frac{\rho_{\text{sat}}}{\rho(T = 0)} &=& \frac{e^2(n_1 \tau_{\text{imp}}^{(1)}/m_1 + n_2 \tau_{\text{imp}}^{(2)}/m_2) }{\frac{e^2(n_2 - n_1)^2}{m_1 n_1 /\tau_{\text{imp}}^{(1)} + m_2 n_2 /\tau_{\text{imp}}^{(2)}}} = \frac{1}{(n_2 - n_1)^2} \left[n_1^2 + n_2^2 + n_1n_2\left(\frac{m_1\tau_{\text{imp}}^{(2)}}{m_2\tau_{\text{imp}}^{(1)}} + \frac{m_2\tau_{\text{imp}}^{(1)}}{m_1\tau_{\text{imp}}^{(2)}} \right)\right].
\end{eqnarray}
This is roughly estimated as $\rho_{\text{sat}}/\rho(T = 0) \sim [(n_1 + n_2)/(n_1 - n_2)]^2$.

Using $\bm{v}_{\text{d}}^{(l)}$, we can explicitly write down the distribution function $\delta f^{(l)}(\bm{k})$~\cite{Lee2021} and we can calculate $L^{(\text{RTA})}_{22}$ as 
\begin{eqnarray}
L^{(\text{RTA})}_{22}
&=& \sum_{l} \frac{n_l}{m_l} \tau^{(l)} (\braket{(\varepsilon_{l,\bm{k}} - \mu)^2}_l -\braket{\varepsilon_{l,\bm{k}} - \mu}_l^2 ) \nonumber \\
&&+ \frac{1}{m_{1}m_{2}}\left(\frac{1}{\tau^{(1)}_{\text{imp}}\tau^{(2)}_{\text{imp}}} + \frac{1}{\tau^{(12)}_{\text{e-h}}\tau^{(2)}_{\text{imp}}} + \frac{1}{\tau^{(21)}_{\text{e-h}}\tau^{(1)}_{\text{imp}}} \right)^{-1} \nonumber \\
&&\times ~\Biggl[n_{1}\braket{\varepsilon_{1,\bm{k}} - \mu}_1 \left(m_{2}\left(\frac{1}{\tau^{(2)}_{\text{imp}}} + \frac{1}{\tau^{(21)}_{\text{e-h}}} \right)\braket{\varepsilon_{1,\bm{k}} - \mu}_1 - \frac{m_{1}}{\tau^{(12)}_{\text{e-h}}} \braket{\varepsilon_{2,\bm{k}} - \mu}_2  \right) \nonumber \\
&& ~~~ + n_{2}\braket{\varepsilon_{2,\bm{k}} - \mu}_2 \left( m_{1}\left(\frac{1}{\tau^{(1)}_{\text{imp}}} + \frac{1}{\tau^{(12)}_{\text{e-h}}} \right)\braket{\varepsilon_{2,\bm{k}} - \mu}_2 - \frac{m_{2}}{\tau^{(21)}_{\text{e-h}}} \braket{\varepsilon_{1,\bm{k}} - \mu}_1  \right) \Biggr],
\end{eqnarray}
where
\begin{eqnarray}
\frac{1}{\tau^{(1)}} &=&  \frac{1}{\tau^{(1)}_{\text{imp}}} + \frac{1}{\tau^{(1)}_{\text{e-e}}} + \frac{1}{\tau^{(12)}_{\text{e-h}}}, \\
\frac{1}{\tau^{(2)}} &=&  \frac{1}{\tau^{(2)}_{\text{imp}}} + \frac{1}{\tau^{(2)}_{\text{e-e}}} + \frac{1}{\tau^{(21)}_{\text{e-h}}}.
\end{eqnarray}
We straightforwardly obtain the thermal conductivity as
\begin{eqnarray}
\kappa^{(\text{RTA})}_{\text{el}} &=& \frac{1}{T}\left( L^{(\text{RTA})}_{22} - \frac{L^{(\text{RTA})}_{21}L^{(\text{RTA})}_{12}}{L^{(\text{RTA})}_{11}}\right) \nonumber \\
&=& \frac{1}{T} \sum_{l}\frac{n_l}{m_l} \tau^{(l)} (\braket{(\varepsilon_{l,\bm{k}} - \mu)^2}_{l} -\braket{\varepsilon_{l,\bm{k}} - \mu}_{l}^2 ) \nonumber \\
&&+ \frac{n_{1}n_{2}}{Tm_{1}m_{2}}  \left[ \braket{\varepsilon_{1,\bm{k}} - \mu }_{1}- \braket{\varepsilon_{2,\bm{k}} - \mu }_{2} \right]^2 \left[  
  \frac{n_{1}}{m_{1}\tau^{(2)}_{\text{imp}}} + \frac{n_{2} - n_{1}}{m_{2}\tau^{(12)}_{\text{e-h}}} + \frac{n_{1} - n_{2}}{m_{1}\tau^{(21)}_{\text{e-h}}} + \frac{n_{2}}{m_{2}\tau^{(1)}_{\text{imp}}}
\right]^{-1}, \label{kappa_RTA}
\end{eqnarray}
where the second term is interpreted as the ambipolar contribution. 

\subsection{Behavior of the ambipolar contribution}
Using eq. (\ref{constraint_RTA}), we can see that the ambipolar contribution in eq. (\ref{kappa_RTA}) is proportional to $(n_1 - n_2)^{-2}$ leading to the suppression of the Lorenz ratio upon doping. In the limit of vanishing of the impurity scattering, the thermal conductivity in the uncompensated case remains finite whereas the thermal conductivity in the compensated case diverges \cite{Zarenia2019_BLG, Lee2021}. Therefore, we expect that the Lorenz ratio becomes zero in this limit when uncompensated since the electrical resistivity becomes zero. Note that the second term in $L^{(\text{RTA})}_{22}$ diverges for both compensated and uncompensated cases when the impurity scattering is absent. This means that the cancellation between the second term in $L^{(\text{RTA})}_{22}$ and $L^{(\text{RTA})}_{21}L^{(\text{RTA})}_{12}/L^{(\text{RTA})}_{11}$, which becomes large in the uncompensated case, is key to removing diverging behavior of $\kappa^{(\text{RTA})}_{\text{el}}$. 

\setcounter{equation}{0}
\setcounter{figure}{0}
\setcounter{table}{0}
\section{Correspondence between the variational method and the RTA}
In this section, we show the correspondence between the variational method and the RTA. The terms $2({}^{t}\bm{J}_1 P_{11}^{-1} \bm{J}_1)$ and $2({}^{t}\bm{J}_1 P_{11}^{-1} \bm{U}_1)$ in the variational method correspond to $L^{(\text{RTA})}_{11}$ and $L^{(\text{RTA})}_{12}$, respectively. Therefore, $\tilde{S} = {}^{t}\bm{J}_1 P_{11}^{-1} \bm{U}_1/T ({}^{t}\bm{J}_1 P_{11}^{-1} \bm{J}_1)$ is equivalent to $S^{(\text{RTA})}$. 
First, $P_{11}^{-1}$ is given by
\begin{eqnarray}
P_{11}^{-1} 
&=& \left(
\begin{array}{cc}
P^{(11)}_{\text{imp},11} + P^{(11)}_{\text{e-h},11} & P^{(12)}_{\text{e-h},11} \\
P^{(21)}_{\text{e-h},11} & P^{(22)}_{\text{imp},11} +  P^{(22)}_{\text{e-h},11}
\end{array}
\right)^{-1} \nonumber \\
&=& \frac{1}{P^{(11)}_{\text{imp},11}P^{(22)}_{\text{imp},11} + P^{(22)}_{\text{e-h},11}P^{(11)}_{\text{imp},11} + P^{(11)}_{\text{e-h},11}P^{(22)}_{\text{imp},11} } \left(
\begin{array}{cc}
P^{(22)}_{\text{imp},11} +  P^{(22)}_{\text{e-h},11} & -P^{(12)}_{\text{e-h},11} \\
-P^{(21)}_{\text{e-h},11} & P^{(11)}_{\text{imp},11} + P^{(11)}_{\text{e-h},11}  
\end{array}
\right),
\end{eqnarray}
where we used a relation of $P^{(11)}_{\text{e-h},11}P^{(22)}_{\text{e-h},11}-P^{(12)}_{\text{e-h},11}P^{(21)}_{\text{e-h},11} = 0$ resulting from eq.~(\ref{constraint_P_11}). Therefore, we obtain 
\begin{eqnarray}
&&{}^{t}\bm{J}_1 P_{11}^{-1} \bm{J}_1 \nonumber \\
&=& \frac{J^{(1)}_1P^{(22)}_{\text{imp},11}J^{(1)}_1 + J^{(2)}_1P^{(11)}_{\text{imp},11}J^{(2)}_1 + J^{(1)}_1P^{(22)}_{\text{e-h},11}J^{(1)}_1 + J^{(2)}_1P^{(11)}_{\text{e-h},11}J^{(2)}_1 - J^{(1)}_1P^{(12)}_{\text{e-h},11}J^{(2)}_1 - J^{(2)}_1P^{(21)}_{\text{e-h},11}J^{(1)}_1}{P^{(11)}_{\text{imp},11}P^{(22)}_{\text{imp},11} + P^{(22)}_{\text{e-h},11}P^{(11)}_{\text{imp},11} + P^{(11)}_{\text{e-h},11}P^{(22)}_{\text{imp},11} } \nonumber \\
&\simeq& \frac{J^{(1)}_1P^{(22)}_{\text{imp},11}J^{(1)}_1 + J^{(2)}_1P^{(11)}_{\text{imp},11}J^{(2)}_1 + (1 - \chi^3)^2 J^{(1)}_1P^{(22)}_{\text{e-h},11}J^{(1)}_1}{P^{(11)}_{\text{imp},11}P^{(22)}_{\text{imp},11} + P^{(22)}_{\text{e-h},11}P^{(11)}_{\text{imp},11} + P^{(11)}_{\text{e-h},11}P^{(22)}_{\text{imp},11} },
\end{eqnarray}
where we used $m_{1} \chi^3 J_{1}^{(1)} \simeq m_{2} J_{1}^{(2)}$ from eqs.~(\ref{J_el_low}) and (\ref{J_ho_low}). The calculation of ${}^{t}\bm{J}_1 P_{11}^{-1} \bm{U}_1 $ is as follows:
\begin{eqnarray}
&&{}^{t}\bm{J}_1 P_{11}^{-1} \bm{U}_1 \nonumber \\
&=& \frac{J^{(1)}_1P^{(22)}_{\text{imp},11}U^{(1)}_1 + J^{(2)}_1P^{(11)}_{\text{imp},11}U^{(2)}_1 + J^{(1)}_1P^{(22)}_{\text{e-h},11}U^{(1)}_1 + J^{(2)}_1P^{(11)}_{\text{e-h},11}U^{(2)}_1 - J^{(1)}_1P^{(12)}_{\text{e-h},11}U^{(2)}_1 - J^{(2)}_1P^{(21)}_{\text{e-h},11}U^{(1)}_1}{P^{(11)}_{\text{imp},11}P^{(22)}_{\text{imp},11} + P^{(22)}_{\text{e-h},11}P^{(11)}_{\text{imp},11} + P^{(11)}_{\text{e-h},11}P^{(22)}_{\text{imp},11} } \nonumber \\
&\simeq& \frac{J^{(1)}_1P^{(22)}_{\text{imp},11}U^{(1)}_1 + J^{(2)}_1P^{(11)}_{\text{imp},11}U^{(2)}_1 + (1 - \chi^3) J^{(1)}_1P^{(22)}_{\text{e-h},11}(U^{(1)}_1 - m_{2}U^{(2)}_1/m_{1} )}{P^{(11)}_{\text{imp},11}P^{(22)}_{\text{imp},11} + P^{(22)}_{\text{e-h},11}P^{(11)}_{\text{imp},11} + P^{(11)}_{\text{e-h},11}P^{(22)}_{\text{imp},11} }.
\end{eqnarray}
In these expressions, we find the correspondences $1/\tau^{(l)}_{\text{imp}} \leftrightarrow eP^{(ll)}_{\text{imp},11}/J^{(l)}_1$ and $1/\tau^{(12)}_{\text{e-h}} \leftrightarrow eP^{(11)}_{\text{e-h},11}/J^{(1)}_1$ ($1/\tau^{(21)}_{\text{e-h}} \leftrightarrow eP^{(22)}_{\text{e-h},11}/J^{(2)}_1$). Therefore, $2({}^{t} \bm{J}_1 P_{11}^{-1}\bm{J}_1)$ and $2({}^{t} \bm{J}_1 P_{11}^{-1}\bm{U}_1)$ are equiavalent to $L^{(\text{RTA})}_{11}$ and $L^{(\text{RTA})}_{12}$. We note that we can also confirm the correspondence between $2({}^{t} \bm{U}_1 P_{11}^{-1}\bm{U}_1)$ and the second term in $L_{22}^{(\text{RTA})}$, which is the ambipolar contribution.

\section{Analysis of $L_{12}$ in the uncompensated case}
We show that the dominant contribution in eq.~(11) comes from $2({}^{t}\bm{J}_1 P_{11}^{-1} \bm{U}_1)$ in the uncompensated case when the electron-hole and intraband scattering dominate over the impurity scattering.
In this case, ${}^{t}\bm{J}_2 P_{22}^{-1} \bm{U}_2$ becomes
\begin{eqnarray}
&&{}^{t}\bm{J}_2 P_{22}^{-1} \bm{U}_2 \nonumber \\
&\simeq & \frac{J^{(1)}_2(P^{(22)}_{\text{e-h},22} + P^{(22)}_{\text{e-e},22} )U^{(1)}_2 + J^{(2)}_2(P^{(11)}_{\text{e-h},22} + P^{(11)}_{\text{e-e},22} )U^{(2)}_2 - J^{(1)}_2P^{(12)}_{\text{e-h},22}U^{(2)}_2 - J^{(2)}_2P^{(21)}_{\text{e-h},22}U^{(1)}_2}
{(P^{(11)}_{\text{e-h},22} + P^{(11)}_{\text{e-e},22} )(P^{(22)}_{\text{e-h},22} + P^{(22)}_{\text{e-e},22} ) - P^{(21)}_{\text{e-h},22}P^{(12)}_{\text{e-h},22} }.
\end{eqnarray}
Note that the quadric terms of $P^{(lk)}_{\text{e-h},22}$ and $P^{(ll)}_{\text{e-e},22}$ in the denominator does not vanish unlike ${}^{t}\bm{J}_1 P_{11}^{-1} \bm{U}_1$.
If we adopt rough order estimations $J^{(l)}_2 \sim (k_BT)^2/\Delta \cdot J^{(l)}_1$, $U^{(l)}_2 \sim \Delta U^{(l)}_1$ and $P^{(lk)}_{22} \sim (k_BT)^2 P^{(lk)}_{11}$, the order of ${}^{t}\bm{J}_2 P_{22}^{-1} \bm{U}_2/{}^{t}\bm{J}_1 P_{11}^{-1} \bm{U}_1$ is estimated as
\begin{eqnarray}
\frac{{}^{t}\bm{J}_2 P_{22}^{-1} \bm{U}_2}{{}^{t}\bm{J}_1 P_{11}^{-1} \bm{U}_1} 
\sim \frac{J^{(1)}_2U^{(1)}_2}{P^{(11)}_{\text{e-h},22 } } 
\times \frac{P^{(11)}_{\text{imp},11} } {(1 - \chi^3) J^{(1)}_1U^{(1)}_1} 
\sim \frac{P^{(11)}_{\text{imp},11}}{(1 - \chi^3) P^{(11)}_{\text{e-h},11}} \ll 1.
\end{eqnarray}
For $2({}^{t}\bm{J}_1 P_{11}^{-1}P_{12}P_{22}^{-1}\bm{U}_2)$ in eq.~(11), we can conduct a similar estimation and show ${}^{t}\bm{J}_1 P_{11}^{-1}P_{12}P_{22}^{-1}\bm{U}_2 \ll {}^{t}\bm{J}_1 P_{11}^{-1} \bm{U}_1$ using $P^{(lk)}_{12} \sim (k_BT)^2/\Delta \cdot  P^{(lk)}_{11}$ and 
\begin{eqnarray}
&& P_{11}^{-1}P_{12} \nonumber \\ 
&=& \frac{1}{P^{(11)}_{\text{imp},11}P^{(22)}_{\text{imp},11} + P^{(22)}_{\text{e-h},11}P^{(11)}_{\text{imp},11} + P^{(11)}_{\text{e-h},11}P^{(22)}_{\text{imp},11} } \nonumber \\
&& \times
\left(
\begin{array}{cc}
P^{(22)}_{\text{imp},11}P^{(11)}_{\text{imp},12} +  P^{(22)}_{\text{e-h},11}P^{(11)}_{\text{imp},12} + P^{(11)}_{\text{e-h},12}P^{(22)}_{\text{imp},11} & P^{(12)}_{\text{e-h},12}P^{(22)}_{\text{imp},11} - P^{(12)}_{\text{e-h},11}P^{(22)}_{\text{imp},12}\\
-P^{(21)}_{\text{e-h},11}P^{(11)}_{\text{imp},12} + P^{(21)}_{\text{e-h},12}P^{(11)}_{\text{imp},11} & P^{(11)}_{\text{imp},11} P^{(22)}_{\text{imp},12} + P^{(11)}_{\text{e-h},11}P^{(22)}_{\text{imp},12}  + P^{(22)}_{\text{e-h},12} P^{(11)}_{\text{imp},11}
\end{array}\right),
\end{eqnarray}
where eqs.~(\ref{constraint_P_11}) $\sim$ (\ref{constraint_P_21_2}) are used.
Therefore, if the electron-hole scattering dominates, $ L_{12} \simeq 2({}^{t}\bm{J}_1 P_{11}^{-1} \bm{U}_1)$ and the Seebeck coefficient of the uncompensated case is evaluated as
\begin{eqnarray}
S 
&\simeq& \frac{1}{T} \frac{(1 - \chi^3) J^{(1)}_1P^{(22)}_{\text{e-h},11}(U^{(1)}_1 - m_{2}U^{(2)}_1/m_{1} )}{(1 - \chi^3)^2 J^{(1)}_1P^{(22)}_{\text{e-h},11}J^{(1)}_1} \nonumber \\
&=& \frac{1}{T} \frac{U^{(1)}_1 - m_{2}U^{(2)}_1/m_{1} }{(1 - \chi^3) J^{(1)}_1} \propto \frac{(m_{1} + \chi m_{2})(\chi^2 m_{1} + m_{2})}{\chi^3 - 1}.
\end{eqnarray}

\bibliographystyle{apsrev4-2}
\bibliography{supplement}